\documentclass[twocolumn,showpacs,preprintnumbers,amsmath,amssymb,superscriptaddress]{revtex4}
\usepackage{graphicx}
\usepackage{dcolumn}
\usepackage{bm}
\def\be{\begin{equation}}
\def\ee{\end{equation}}
\def\br{\begin{eqnarray}}
\def\er{\end{eqnarray}}
\def\brn{\begin{eqnarray*}}
\def\ern{\end{eqnarray*}}

\begin{document}

\title{Quasi-free (p,2p) and (p,pn)  reactions with unstable nuclei}
\author{T. Aumann}
\email[E-mail:~]{t.aumann@gsi.de}
\affiliation{Institut f\"ur Kernphysik, Technische Universit\"at Darmstadt, Schlossgartenstr. 9, 64289 Darmstadt, Germany}
\affiliation{GSI Helmholtzzentrum f\"ur Schwerionenforschung, Planckstr. 1, 64291 Darmstadt, Germany}
\author{ C.A. Bertulani}
\email[E-mail:~]{carlos.bertulani@tamuc.edu}
\affiliation{Texas A\&M University-Commerce, PO Box 3011, Commerce, TX 75429, USA}
\author{J. Ryckebusch}
\email[E-mail:~]{Jan.Ryckebusch@ugent.be}
\affiliation{Department of Physics and Astronomy, Ghent University, B-9000 Gent, Belgium}

\begin{abstract}
We study (p,2p) and (p,pn) reactions at proton energies in the range of 100 MeV -- 1 GeV. Our purpose is to explore the most sensitive observables in unpolarized reactions with inverse kinematics involving radioactive nuclei. We formulate a model based on the eikonal theory to describe total cross sections and momentum distributions of the recoiled residual nucleus. The model is similar to the one adopted for knockout reactions with heavy ions. We show that momentum distributions are sensitive to the angular momentum of the ejected nucleon which can be used as an spectroscopic tool. The total cross sections are sensitive to the nucleon separation energies and to  multiple scattering effects. Our calculations also indicate that a beam energy around 500 MeV/nucleon has a smaller dependence on the anisotropy of the nucleon-nucleon elastic scattering.   
\end{abstract}

\pacs{21.00.00, 25.55.Ci}

\maketitle

\section{Introduction}

Quasifree (p, pN) reactions (N=proton or neutron) represent one of the most 
common experimental tools to access information on single-particle properties
in nuclei. 
In quasifree (p, pN) scattering an incident proton of
medium energy (typically several hundred MeV) knocks out a bound
nucleon. The energy spectrum of the outgoing nucleons provides information on the energy and other quantum numbers of the struck nucleon in the nucleus. The shape of the angular correlations of the outgoing nucleons, or the recoil momentum of the nucleus, is connected to the momentum distribution of the knocked-out nucleon.  In the past four decades
quasifree-scattering experiments have been performed
with this basic purpose. 
For seminal  reviews on (p,pN) reactions
see, e.g., Refs. \cite{JM73,Kit85}.

The theory developments in (p,2p) reactions have been largely done in the past 50 years. Very few theorists still work in this field, basically due to 
the decrease in the number of experiments carried out in such a fashion. However, the availability of high-energy radioactive beams allows in principle 
to utilize the method of quasifree scattering in inverse kinematics with hydrogen targets. This will open huge possibilities to investigate properties of unstable
nuclei  such as systematic studies of single-particle structure or nucleon-nucleon correlations as a function of neutron-to-proton asymmetry. 

So far, knockout reactions using composite targets have been used extensively to investigate the 
shell structure of rare isotopes and many valuable
results have been obtained \cite{HT03,GA08}.  %A. Gade et al., PRC 77, 044306 (2008)
Gade et al.  \cite{GA08} have discussed, for instance, the reduction of spectroscopic strength in dependence on the asymmetry of the neutron and proton 
Fermi energies deduced from measured knockout cross sections. Theoretically, such an effect is indeed expected both for nuclei \cite{BD09, JE11}
and asymmetric nuclear matter  \cite{FR05}; the predicted effects, however, cannot explain the data. 

Quasifree scattering in inverse kinematics
potentially provides a more sensitive tool to study such effects since the reaction is less surface dominated. The possibility to detect all outgoing
particles can provide a kinematically complete measurement of the reaction. 
At GSI, RIKEN, and other laboratories, experiments are being planned to study such effects more systematically and by applying 
a kinematically complete measurement of quasi-free knockout reactions \cite{r3b, AU07, Kob08},
and first pilot experiments were successful \cite{Kob08, PA12}. 
In order to extract the physics observables from the measured (p,2p) and (p,pn) cross sections, reaction theory plays a key role.
However, much of the theoretical expertise in this field was lost. 
It is thus imperative for the community to concentrate theoretical efforts in this problem with aim at the upcoming experiments with radioactive beams. 

The framework of the distorted wave impulse approximation (DWIA) is often used in numerical calculations of (p,pN) reactions \cite{CR77,CR83}. 
This method assumes that the dominant mechanism for the knock\-out reaction  is due to a single interaction  between  the incident particle and the 
struck nucleon. The effect of the coherent multiple scattering with the other nucleons is incorporated by using  distorted waves calculated from a mean
nuclear potential, including absorption due to excitations to other channels. One also needs to account for the medium modification of interaction between 
the incoming proton and the struck nucleon \cite{KMT59,M3Y,JLM,LF81,YIM83,TY85,FL85,HI86,NL88}.

A part of the difficulties in studying (p,2p) and (p,pn) reactions relies on the uncertainty of in-medium interactions, and ambiguities from the reaction mechanism. 
Even in the energy region of several hundred MeV, where the NN cross section shows its minimum and the reaction mechanism is expected to be simplest,
 multi-step processes are not negligible in general \cite{Ma58,JMS76,KMRV95,Ove06,Ove07,CR09}.
 
Measurements of polarization observables in quasi elastic (p,2p) reaction at intermediate energies give an even better information on nuclear shell structure and
multiple scattering effects, and also on possible modification of the nucleon\-nucleon (NN) interaction parameters in nuclear medium \cite{Ma58,JMS76}. Due to
the nuclear spin-orbit coupling and absorption of the projectile and secondary protons in the nuclear matter the proton knocked out from a particular nuclear 
shell with orbital moment $l\ne 0$ can be polarized \cite{JM73,Kit85,Ove07,CR09,RCV11,VRC12}. 

In the next sections we describe a formalism for quasi-free knockout reactions, first by describing the state-of-the-art DWIA method which has been shown to reproduce experimental data with a good accuracy \cite{JM73,Kit85}. Then we introduce a few simplifications which can be justified with the use of eikonal scattering waves, 
appropriate for high energy scattering. Finally, we apply the formalism to a detailed analysis of several observables which can be assessed in  spectroscopic studies of rare nuclear isotopes probed with (p,pN) reactions in inverse kinematics.

\section{Quasi-free scattering formalism}

\subsection{Distorted wave impulse approximation}

The standard DWIA expression for the quasi-free cross
section is \cite{JM73,Kit85}
\begin{equation}
{d^3\sigma \over dT_N d\Omega^\prime_p d\Omega_N} =K^\prime {d\sigma_{pN}\over d\Omega} |F({\bf Q})|^2,
\label{int}
\end{equation}
where $K^\prime$ is a kinematic factor, $|F({\bf Q})|^2$ is the momentum distribution of the knocked-out nucleon $N$ in the nucleus and $d\sigma_{pN}/d\Omega$ is the free, or quasi-free, pN cross section. In this formalism the off-shell pN t-matrix is required, and the factorized form that appears in Eq.~\eqref{int}
is valid only if off-shell effects are not very important. In proton-induced knock-out
reactions at high energies this hypothesis has been confirmed  in previous works  \cite{JM73,Kit85}. 
\subsection{Transition amplitude}
In the DWIA, the transition amplitude for
the A(p, pN)B reaction is given by \cite{JM73}
\begin{equation}
T_{p,pN}=\sqrt{S(lj)}\left< \chi_{{\bf k}_p^\prime}^{(-)} \chi_{{\bf k}_N}^{(-)}          
\left| \tau_{pN}\right| \chi_{{\bf k}_p}^{(+)}\psi_{jlm}\right>\label{Tmat}
\end{equation}
where $\chi_{{\bf k}_p^\prime}^{(-)}$ ($\chi_{{\bf k}_N}^{(-)}$) is the distorted wave for an outgoing proton (knocked-out nucleon) in the presence of the
residual nucleus $B$, $\chi_{{\bf k}_p}^{(+)}$ is the distorted wave for an incoming proton in the presence of the target nucleus $A$, and $ \psi_{jlm}$ is the bound state wave function of the knocked-out nucleon;
$ \sqrt{S(lj)}$ is the corresponding spectroscopic amplitude for a bound nucleon with quantum numbers ($lj$). 
Later, we will define the energy $E$ at which the two-body pN scattering matrix $\tau_{pN}$ is evaluated. 

\begin{figure}
[ptb]
\begin{center}
\includegraphics[
width=3.5in
]%
{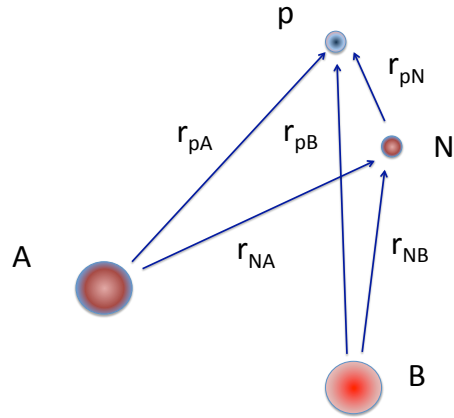}%
\caption{(Color online). The coordinates used in the text are shown. }%
\label{coord}%
\end{center}
\end{figure}

In coordinate space the matrix element given by Eq.~\eqref{Tmat} can be written as
\begin{eqnarray}
T_{p,pN}&=&\sqrt{S(lj)} \int d^3{\bf r}^\prime_{pB} d^3{\bf r}^\prime_{NB}
d^3{\bf r}_{pA} d^3{\bf r}_{NB}   \nonumber \\
&\times& \tau  ({\bf r}_{pB},{\bf r}_{NB};{\bf r}^\prime_{pA}, {\bf r}_{NB}^\prime )
\nonumber \\
&\times&\chi_{{\bf k}_p^\prime}^{(-) *} ({\bf r}^\prime_{pB}) \chi_{{\bf k}_N}^{(-) *}({\bf r}^\prime_{NB}) \chi_{{\bf k}_p}^{(+)} ({\bf r}_{pA})\psi_{jlm}({\bf r}_{NB}),
 \nonumber \\
\label{tmat2}
\end{eqnarray}
where the scattering waves are normalized so that $$\int d^3{\bf r} \chi^*_{\bf k}({\bf r}) \chi_{\bf k^\prime}({\bf r}) =  \delta({\bf k}-{\bf k^\prime}),$$ and the bound-state wavefunction $\psi_{nlj}$ is normalized to the unity, $$\int d^3{\bf r} |\psi_{nlj}({\bf r})|^2=1.$$ 

An inspection of the integrand in Eq. \eqref{tmat2} can help us to eliminate several of the integrals. The coordinates of the proton and the nucleon $N$ are related through ${\bf r}_{pA}={\bf r}_{pN}+ {\bf r}_{NA}$. Since the range of the
pN interaction is much smaller than the nuclear size, the integral in Eq. \eqref{tmat2} will sample small values of $r_{pN}$ such that $r_{pN} \ll r_{NA}$. 
The T-matix in Eq. \eqref{tmat2} reduces to an integral only over  the ${\bf r}_{NB}$ coordinate  \cite{JM73,Kit85},
\begin{eqnarray}
T_{p,pN}&=&\sqrt{S(lj)}\tau  ({\bf k}^\prime_{pN},{\bf k}_{pN};E) \int d^3{\bf r}_{NB} \nonumber \\&\times&\chi_{{\bf k}_p^\prime}^{(-) *} ({\bf r}_{NB}) \chi_{{\bf k}_N}^{(-) *}({\bf r}_{NB}) \chi_{{\bf k}_p}^{(+)} (\alpha {\bf r}_{NB})\psi_{jlm}({\bf r}_{NB}),\nonumber \\
\label{tmat5}
\end{eqnarray}
where $\alpha = (A-1)/A$ and $\tau({\bf k}^\prime_{pN},{\bf k}_{pN};E)$ is the Fourier transform of the pN t-matrix in Eq. \eqref{tmat2}.

The equation above  has been the pillar of nearly all numerical calculations of quasi-free (p,2p) reactions. 
The factorization of the matrix element in Eq. \eqref{tmat5} is exact in plane wave impulse approximation (PWIA).
The argument used to justify it in DWIA is that the two-body
interaction is of sufficiently short-range so that the distorted
waves do not change significantly over the range which
contributes significantly to the matrix element \eqref{tmat2}.
Corrections to this approximation have been studied by several authors (see, e.g. \cite{Dap82}) and the results have been inconclusive in view of the several other approximations that have been further introduced.

Perhaps, the most serious approximation used to obtain Eq.~\eqref{tmat5} is the assumption of quasi-free binary scattering, i.e. without the effects of multiple collisions, and other many-body effects such as antisymmetrization. The corrections due to these effects have been also reported in the literature, e.g. in Refs.~\cite{Jai79,Dap82}. The conclusions are that the deviations from the impulse approximation might be important, but they are also subject to uncertainty due to the use of several other approximations. We will thus retain the DWIA as our tool of choice, although we consider other corrections below. It should be emphasized that nonlocality corrections and spin-orbit terms in the optical potentials might play a relevant role and are not considered here. In Ref. \cite{Sam86}  it was shown that the factorization approximation is valid for proton energies as low as 75 MeV. We will consider much higher energies here for which more simplifications can be done.

\subsection{Cross section}

The differential cross section is given by \cite{Dap76}
\begin{equation}
{d^3\sigma \over dE_p^\prime d\Omega_p^\prime d\Omega_N} ={K \over (2s_p+1)(2J_A+1)}
\sum_\gamma \left| T_{p,pN}(\gamma)\right|^2
\label{cross1}
\end{equation}
where $E_p^\prime$ is the proton energy in the final channel, and the kinematical factor $K$ is  given by (here we use $k_i=p_i/\hbar$) \cite{Mar58}
\begin{eqnarray}
K&=&{m_p^2m_N c^6\over (\hbar c)^6 (2\pi )^5} {k_p^\prime k_N\over k_p}\nonumber \\
&\times& \left| 1 + {m_N \over M_B} \left[1- {k_p\over k_N}\cos \theta_N +{k_p^\prime \over k_N}
\cos(\theta_p +\theta_N)\right]\right|^{-1} \ .
\label{cross2}
\end{eqnarray}

We define a {\it missing momentum}, ${\bf p}_m$, a {\it missing energy}, $\epsilon_m$,  and a {\it missing mass}, $m_m$,  by means of
 \begin{eqnarray}
\epsilon_m&=&E_N+ E_{p}^\prime- E_p-m_N\nonumber\\
 {\bf p}_m &=&{\bf k}_N+{\bf k}_{p}^\prime - {\bf k}_p \nonumber \\
 m_m^2&=&\epsilon_m^2-p_m^2. 
\label{cross2b}
\end{eqnarray}
The removal of the nucleon with momentum, $p_m$, from the nucleus implies a transfer of that three momentum from the nucleus to the observed proton and knockout nucleon final state. For an exclusive reaction, only a
small deficit in the final-state energy, $\epsilon_m$ can be observed.

The spin (projection) quantum numbers of the particle $p$ and the target nucleus $A$ in the initial state are $s_p$($\mu_p$) and $J_A$($M_A$), respectively. Also, in the final state the quantum numbers of the particles $p$ and $N$ and the residual nucleus $B$ are $s_p$($\mu_p^\prime$), $s_N$($\mu_N$), and $J_B$($M_B$), respectively. The summation $(\gamma$) of the transition matrix $T_{p,pN}$ in Eq.~\eqref{cross1} is taken over the spin components 
$$\gamma = (\mu_p, \mu_p^\prime, \mu_N, M_A, M_B),$$ 
in the initial and final states.
In the spin summation, the transition matrices will have an explicit dependence on 
$s_p$($\mu_p$, $\mu_p^\prime$), $s_N$($\mu_N$), $J_A$($M_A$) and $J_B$($M_B$). Except for specific cases, we will use the simpler notation $jlm$ for the angular momentum quantum numbers of the nucleons.

\subsection{Plane wave impulse approximation}
Physical insight is obtained by the use of the PWIA approximation, i.e., no scattering wave distortion. The relation between the pN scattering amplitude (in the pN c.m.) and the pN t-matrix is (see, e.g., \cite{HRB91})
\begin{equation}
f_{pN}(\theta ; E)=-{2\pi^2 m\over \hbar^2}\tau({\bf k}^\prime_{pN},{\bf k}_{pN};E), \label{ftheta}
\end{equation}
in terms of which the (elastic) scattering cross section in the pN c.m. system is 
\begin{equation}
{d\sigma_{pN}\over d\Omega} = \left|f_{pN}(\theta; E)\right|^2 .
\end{equation}

If the wavefunctions in Eq. \eqref{tmat5} are replaced by plane-waves, one gets
\begin{eqnarray}
T_{p,pN}^{(PWIA)}=\sqrt{S(lj)}\tau({\bf k}^\prime_{pN},{\bf k}_{pN};E) \int d^3{\bf r} \ e^{-i{\bf Q.r}}\psi_{jlm}({\bf r}),\nonumber \\
\label{cross5}
\end{eqnarray}
where ${\bf Q}$ is the missing momentum defined  in Eq.~\eqref{cross2b}, and modified to
\begin{equation}
{\bf Q}= {\bf k}_p^\prime+{\bf k}_N-\alpha{\bf k}_p ,
\label{cross6}
\end{equation}
where we introduced a correction $\alpha = (A-1)/A$ to account for c.m. motion \cite{JM73}.
Thus, we reproduce Eq.~\eqref{int}, with 
\begin{equation}
F({\bf Q})= \int d^3{\bf r} \ e^{-i{\bf Q.r}}\psi_{jlm}({\bf r}) \label{ftpsi}
\end{equation} 
and the kinematic factor in Eq.~\eqref{Tmat} is given in terms of the  kinematic factor of Eq. \eqref{cross1} as
\begin{equation}
K^\prime=\left({\hbar^2\over 2\pi^2 m}\right)^2 \ K .
\end{equation}
Fig. \ref{coord}  show the coordinates used in the text. 

%In terms of $\theta_p^{\prime}$ and $\theta_N$, the momentum transfer in Eq. \eqref{cross6} is given by
%\begin{eqnarray}
%Q^2&=&k_{p'}^2+k_N^2+\alpha^2k_p^2+2k_pk_N\cos(\theta_p-\theta_N) \nonumber \\
%&-&2\alpha k_pk_{p'} \cos\theta_p - 2 \alpha k_p k_N \cos\theta_N,\nonumber\\
%Q_z&=&k_{p}^{\prime}\cos \theta_p^{\prime}+k_N \cos\theta_N-\alpha k_p,\nonumber\\
%Q_t&=&k_{p}^{\prime} \sin\theta_p^{\prime}+k_N \sin\theta_N, \nonumber \\
%Q^2 & = & Q_{z}^{2} + Q_{t}^{2} ,
%\label{QQQ}
%\end{eqnarray}
%where $Q_z$ ($Q_t$) are the longitudinal (transverse) momentum transfers.    

Eq.~\eqref{cross5} is revealing: it tell us that in the lowest approximation, the (p,2p) and (p,pn) cross sections are proportional to the
momentum distribution of nucleons inside the target, determined by their wavefunction $\psi_{jlm}({\bf r})$. This feature is common in many
direct reactions as has been used in the identification and interpretation of many remarkable phenomena. For example, it
has been used to identify ``halo" structure in exotic nuclei \cite{Tan85}, in the interpretation of momentum distributions in knockout reactions \cite{HM85,BM92,Orr92,HBE96,Au00,Par00,Oza01,HT03,Cor04,BH04,BG06,Kan09,GT10,Ro11}, 
or in the experimental analysis of transfer reactions, such as in the Trojan Horse method for nuclear astrophysics \cite{Bau96,Spi01,Spi03,TB03,Spi04,Cog07,Piz13}.

Evidently, the PWIA is not a good approximation. In fact, distortions due to  absorption and elastic scattering accounted for in DWIA  lead to a deviation from Eq.~\eqref{cross5} and its simple interpretation. But it is still useful for physical understanding  of most results. Although  (p,2p) reactions have been carried out with high-energy protons, only rather recently, eikonal waves, and the Glauber treatment of multiple scattering have been used to account for distortions and absorption \cite{Ove06,Ove07,CR09,RCV11,VRC12}.  One probable reason is that measurements have been carried out for large angle scattering and sometimes large energy transfer, conditions that invalidate the use of the eikonal approximation, although attempts along these lines have been tried already at very early studies of (p,2p) reactions \cite{Mc64,Mc66}.
We will show that the use of eikonal waves is still justified if care is taken to separate incoming and outgoing channels. The advantage of the eikonal formalism in comparison with traditional DWIA formalism as described in Refs.~\cite{JM73,Kit85} is enormous because it allows a much easier treatment of the multiple-scattering problem. 

\subsection{Distorted waves}

The PWIA approximation in Eq.~\eqref{cross5} neglects  important absorption and refraction properties of the  nucleon scattering waves.  In high-energy collisions, a much better result can be obtained by using the eikonal approximation
\begin{eqnarray}
& & \chi_i({\bf r})^{in(out)}  =  \exp \left[i{\bf k_i^{\rm in(out)}\cdot r}\right]
\nonumber \\
& & \times \exp \left[- {i\over \hbar v} \int_{a_{in(out)}}^{b_{in(out)}} dz^\prime U_i^{\rm in(out)}({\bf r}^\prime) \right],
\label{cross6b}
\end{eqnarray}
where ${\bf r}\equiv({\bf b},z)$, $v$ is the velocity of the nucleon $i$ and $U_i$ is the optical potential accounting for all interactions of the particle with the nucleus. The integration limits are $a_{in(out)}=\mp \infty$ and 
the location where the collision occurs inside the projectile. An average of this position is done for each ``impact parameter" $b$.

For practical purposes, one can also write \eqref{cross6b} as
\begin{equation}
\chi_i({\bf r})^{in(out)}={\cal S}_{in(out)}(b)\exp\left[i{\bf k_i^{\rm in(out)}\cdot r}\right], 
\label{cross6c}
\end{equation}
with
\begin{equation} 
{\cal S}_{in(out)}(b)= \exp\left[- {i\over \hbar v} \int_{a_{in(out)}}^{b_{in(out)}} dz^\prime U_i^{\rm in(out)}({\bf r}^\prime)\right] .
\end{equation}
We can interpret ${\cal S}_{in(out)}(b)$ as the ``{\it survival amplitudes}" for the incoming and outgoing waves. They measure the distortion and absorption of the incoming proton and outgoing nucleons as a function of their position in space.

For the real part of the optical potential, it is appropriate to use a microscopic folding potential, such as the M3Y \cite{M3Y} 
potential. The imaginary part of the optical potential, corresponding to reaction loss to other channels, needs to be introduced separately. For high energy collisions, this can be done by assuming that absorption is due to incoherent binary nucleon-nucleon collisions in the t-$\rho\rho$ approximation \cite{HRB91}.  
Then the optical potential entering Eq. \eqref{cross6b} reads 
\begin{eqnarray}
U_i({\bf r})&=& U^{M3Y}_i({\bf r}) +U_i^{(c)}({\bf r})\nonumber \\
&-& i {E_i\over k_i} \sigma_i (E_i) \int \rho_{_{A(B)}}({\bf r}-{\bf r}^\prime) \rho_p({\bf r}^\prime) d^3{\bf r}^\prime , \label{uibfr}
\end{eqnarray}
where $U_i^{(c)}$ is the Coulomb potential between the nucleon $i$ and the nucleus A (entrance channel) or B = A-1 (outgoing channel). As the eikonal integral for the Coulomb field in Eq.~\eqref{cross6b} diverges, a regularized Coulomb phase is used \cite{BG06}.
The last term of Eq.~\eqref{uibfr} relates the imaginary part of the optical potential to the nucleon-nucleon cross section, which depends on the incoming proton, or outgoing nucleon, energy $E_i$. The integral contains the density of the nuclei A or B folded with the nucleon density. 
The intrinsic matter density  of the proton (or neutron), $\rho_p(r)$,   is taken as an exponential
function, corresponding to a form factor $ \rho_p(q) = (1 + q^2/a^2)^{-1}$. We will use $a^2 = 0.71$ fm$^{-2}$, for a proton rms radius of 0.87 fm. The Fourier transforms for the nuclei $A$ and $B$ are obtained from theoretical nuclear densities, calculated by using Hartree-Fock+BCS theory according to Refs. \cite{BLS09,BLS12}.

\subsection{Nucleon-nucleon cross sections}

The free (total) nucleon-nucleon cross sections for $E>10$ MeV were fitted to the data of the
Particle Data Group  \cite{pdgxnn}. The fits are given by the expressions (1) and (2) of Ref.~\cite{BC10}. Their analytical fit reproduces very well the 
experimental total pp and pn cross sections in the energy interval $10\ {\rm MeV} \leq E_{lab} \leq 5\ {\rm GeV}$. 

Medium corrections of the nucleon-nucleon cross sections can be relevant even at moderately high energies ($E_p \sim 200$ MeV). At higher energies, nuclear transparencies can be nicely reproduced in Glauber calculations with free NN cross sections - very small medium corrections cannot be excluded, but the measured transparencies seem to indicate that they are small.  In (p,pN) reactions the final proton or the knocked-out nucleon can have much lower energies than the incident proton, thus raising the importance of medium corrections. 
These corrections for nucleon-nucleon scattering in nuclear matter have been studied in, e.g.,  Refs.~\cite{CV56,Ber86,LM:1993,LM:1994,Xian98,Ber01,BC10,CSB13}. A calculation of medium effects using Brueckner-Hartree-Fock method shows that medium corrections are mainly due to Pauli blocking, which can be linked to functions of the local nucleon densities \cite{CSB13}. The analytical formulation developed in Ref.~\cite{CV56,Ber86,BC10} yields the following parametrization for the nucleon-nucleus cross section in terms of the local nuclear density $\rho$ (in fm$^{-3}$),
\begin{eqnarray}
&&\sigma_{pN}(E_i,\rho)=\sigma_{pN}^{(free)}(E_i)\Bigg[1-{50.12 \rho^{2/3}\over E_i}\nonumber\\
&+&\Theta\left({125.3\rho^{2/3}\over E_i}-1\right){50.12 \rho^{2/3}\over E_i}\left(1-{E_i\over 125.3\rho^{2/3}}\right)^{5/2}\Bigg]\nonumber\\
\label{VM1}
\end{eqnarray}
where $\sigma_{pN}(E_i)$ is the free cross section, $E_i$ is the nucleon laboratory energy in MeV and $\Theta$ is the step function, i.e., $\Theta(x)=0$ if $x<0$, and $\Theta=1$ if $x>0$.  
For an impact parameter $b$ the above cross section is averaged along the incoming (outgoing) longitudinal coordinate $z$. This procedure yields the values of $\overline{\sigma_{pn}}$ and $\overline{\sigma_{pp}}$ to be used in Eq.~\eqref{uibfr}. Further, the cross sections are averaged over the local number of protons and of neutrons (local isospin average). The nucleon-nucleon cross sections entering these expressions are given by
\begin{eqnarray}
\overline{\sigma_{p}}&=&{N_A\over A_A} \overline{\sigma_{pn}} + {Z_A\over A_A}\overline{\sigma_{pp}}, \ \ \ \
\overline{\sigma_{p'}}={N_B \over A_B}\overline{\sigma_{pn}}+{Z_B\over A_B}\overline{\sigma_{pp}} \nonumber \\
&{\rm and}&\overline{\sigma_{n}}={N_B \over A_B}\overline{\sigma_{pp}}+{Z_B\over A_B}\overline{\sigma_{pn}} ,
\label{isoave}
\end{eqnarray}
where ($N_A,Z_A,A_A$) and ($N_B,Z_B,A_B$) are the neutron, charge, and mass numbers of the target and residual nucleus, respectively.  A similar isospin average is used to obtain $\overline{\sigma_N}$, depending if the outgoing nucleon is N=neutron or N=proton.  

\section{Momentum distributions}

\subsection{S-matrices}

Eq.~\eqref{cross6b} allows a simple interpretation of the transition matrix for the scattering of a high energy projectile by a proton in a (p,pN) reaction (N = p, or N = n, for proton or neutron knockout). We adopt a similar approach as in Refs.~\cite{HM85,BM92,HBE96} for knockout reactions in nucleus-nucleus collisions. The two-body wavefunction for the incoming proton-nucleus channel is 
\begin{equation}
\Psi_{in} = {\cal S}_{in} \exp(i\alpha{\bf k}_p \cdot {\bf r}) \psi_{jlm},
\end{equation} 
where ${\cal S}_{in}$ is the scattering matrix (or survival amplitude)  of the incoming proton up to the collision point and $\psi_{jlm}$ is the single-particle wavefunction of the bound nucleon in its initial state. Analogously, the two-body wavefunction for the outgoing channel is given by 
\begin{equation}
\Psi_{out} = {\cal S}_{out}^{(p)} S_{out}^{(N)}\exp[i({\bf k}_N +{\bf k}_{p} ^{\prime}) \cdot {\bf r}],
\end{equation}
where $S_{out}^i$ is the scattering matrices (survival amplitudes) for the outgoing nucleon $i = $ p (proton) or N (knocked-out nucleon). The scattering matrices (survival amplitudes) account for the distortions in the incoming and outgoing channels, as well as for the absorption in those channels. Absorption means that if other binary collisions occur, other channels will open and the contribution to the (p,pN) channel will be lost. Absorption is taken care by the last term of Eq.~\eqref{uibfr}.  Note that in separating the scattering into an incoming and outgoing wave, the conditions of validity of the eikonal approximation, Eq. \ref{cross6b}, are satisfied, as long as the deviations from a straight-line during the incoming and, separately, during the outgoing channels, are kept. The energy loss along these paths in both channels are also expected to be small, the main energy transfer occurring during the quasi-free  p,pN scattering.

The interpretation of the (p,pN) reaction is now straightforward. The transition matrix  is simply given by $$ T_{p,pN}^{(eik)}=\sqrt{S(lj)}\ \tau({\bf k}^\prime_{pN},{\bf k}_{pN};E)  \left\langle \Psi_f | \Psi_i \right\rangle$$ where $\sqrt{S(lj)}$ is the amplitude to find the bound nucleon in the orbital $jlm$, $\tau({\bf k}^\prime_{pN},{\bf k}_{pN};E)$ is the proton-nucleon scattering amplitude, and $ \left\langle \Psi_f | \Psi_i \right\rangle$ are overlap integrals of the initial and final states of the nuclei $A\rightarrow B$. In
an independent particle model with a spectator nucleus B one can write that 
$ \left\langle \Psi_f | \Psi_i \right\rangle \approx \left\langle \Psi_{out} | \Psi_A \right\rangle$. Apart
from kinematical factors, the total scattering amplitude is the product of the free nucleon-nucleon
scattering amplitude times the probability amplitude for finding inside the nucleus a nucleon at position
${\bf r}$.
The equation above can also be rewritten as
\begin{eqnarray}
T_{p,pN}^{(eik)}&=&\sqrt{S(lj)}\tau({\bf k}^\prime_{pN},{\bf k}_{pN};E) \nonumber \\
&\times&\int d^3{\bf r} \ e^{-i{\bf Q.r}}{\cal S}(b,\theta)
\psi_{jlm}({\bf r}),
\label{cross7}
\end{eqnarray}
where ${\bf Q}$ is given by Eq.~\eqref{cross6}, $\theta\equiv \theta(\theta_p^{\prime},\theta_N)$ is a function of the angles $\theta_p^{\prime}$ and $\theta_N$, ans ${\cal S}(b,\theta)$ is the product of scattering matrices for $pA$, $p^\prime B$ and $NB$ scattering, i.e.,
\begin{eqnarray}
{\cal S}(b,\theta)={\cal S}_{pA}(E_p,b)
{\cal S}_{p'B}(E_{p}^{\prime}, \theta_{p} ^{\prime}, b)
{\cal S}_{NB}(E_N,\theta_N,b), \nonumber \\
\label{cross8}
\end{eqnarray}
with the scattering matrices for the initial proton-target $(pA)$, the final proton-residual nucleus $(pB)$, and the nucleon-residual nucleus $(NB)$
scattering.

The semi-inclusive differential scattering cross section for the reaction A(p,pN)B by removing a nucleon from the orbital $jl$ under conditions of fixed ${\bf Q}$ is given by
\begin{eqnarray}
{d\sigma}= {d^3Q \over (2\pi)^3} {1\over 2j+1} \left(-{2\pi^2 m_p\over \hbar^2}\right)^2\sum_m \left| T_{p,pN}^{(eik)} \right|^2 \ ,
\label{dshos}
\end{eqnarray}
where $d^3{\bf Q} / (2\pi)^3$ is the density of  states and the sum takes care of the average over all magnetic substates of the bound-state wavefunction $\psi_{jlm}$. In the last equality, we have made use of the Eq.~(\ref{ftheta}).

We perform an average of $d\sigma_{pN}/ d\Omega$ and the scattering matrices over all possible energies of the final proton and nucleon which lead to the momentum transfer $Q$. Hence, our differential cross section in the Distorted Wave Impulse Approximation (DWIA) is given by
\begin{eqnarray}
\left( {d\sigma\over d ^3 Q}\right)_{DWIA} &=&  {1 \over (2\pi)^3} {S(lj)\over 2j+1} \sum_m \left< {d\sigma_{pN}\over d\Omega} \right>_Q \nonumber\\
&\times& \left| \int d^3 {\bf r} \ e^{-i{\bf Q.r}}\left<{\cal S}(b)\right>_Q
\psi_{jlm}({\bf r}) \right|^2 . \label{momdis}
\end{eqnarray}
Here the S-matrix is also averaged  over all pp$^{\prime}$ scattering angles leading to the same magnitude of the momentum transfer $Q$. 
Eq.~\eqref{momdis} is our starting formula for the calculations of the momentum distributions of the recoiled fragments.

We
neglect the fact that the momenta of the
quasi-free matrix element do not occur in the free
scattering because of the difference in energy conservation
between the two cases. This difference is caused
by the nonzero value of the separation energy in the
quasi-free scattering and by the energy and momentum carried by 
the recoiling nucleus.  In other words,
${d\sigma_{pN}/ d\Omega}$ in Eq. \eqref{dshos} is a half-off-the-energy-shell cross section. In non-relativistic terms, the off-shellness is because energy conservation requires
that $-S_N + p_m^2/2M_B$, where $p_m$  is the missing momentum, or recoil momentum of nucleus $B=A-1$ given by Eq. \eqref{cross2b} and $S_N$ is the nucleon separation energy.  Due to this mismatch, there is a certain arbitrariness in chosen the value of $d\sigma_{pN}/ d\Omega$ in Eq. \eqref{int}. This is an amount $S_N+p_m^2(1/m-1/M_B)/2$ off shell.
The importance of off-energy-shell effects has been investigated in Refs.~\cite{RS70,Cha77,Roo77}. The conclusion
is that off-energy-shell effects are small and can be neglected. Thus, the standard prescription is to
to replace ${d\sigma_{pN}/ d\Omega}$ by the measured on-energy-shell cross section, i.e., the elastic
pN scattering. This approximation amounts only
to a few percent uncertainty as long as one considers quasi-free
processes for proton energies greater than 200 MeV \cite{BM61,JM66}.

The  free $d\sigma_{pN}/ d\Omega$ cross section shows a considerable anisotropy in the laboratory energy range of $E_p =200 -1000$ MeV. Phase-shift analyses of pp  and pn elastic scattering cross sections are available (e.g., Refs. \cite{SKR93,AW94}), yielding elaborate parametrizations that are generally in good agreement with the data \cite{NDB}. We take into account the angular anisotropy and the parametrizations by using a fit to the experimental differential nucleon-nucleon cross sections as explained in Ref. \cite{NDB}.  The angular distributions  were obtained using the computer code SAID, which implements a comprehensive phase-shift analysis that encompasses the world elastic pp and pn-scattering data \cite{AW94}. 

\subsection{Single particle wavefunctions}

If the nuclear states are assumed to be independent on the total angular momentum of the residual nucleus, one can write  
\begin{equation}
\psi_{jlm}  =\frac{u_{lj}\left(  r\right)  }%
{r}\sum_{m_{l} \ m_{s}}\left< lm_l sm_s | jm\right>Y_{lm_{l}}\left(  \widehat{\mathbf{r}}\right)
\chi_{m_s} , \label{psi_expansion2}%
\end{equation}
where $\left< lm_l sm_s | jm\right>$ are Clebsh-Gordan coefficients, $Y_{lm_l}$ are spherical harmonics, and $\chi_{m_s}$ are spinors. The radial wavefunction is normalized so that
$
{\int
%\limits_{0}^{\infty}
}
dr\left\vert u_{lj}\left(  r\right)  \right\vert ^{2}=1. \label{norm}%
$
These wavefunctions are calculated using a Woods-Saxon potential with central, surface, and spin-orbit parts (see, e.g., \cite{BG06}). The integral in Eq.~\eqref{momdis} has cylindrical symmetry which can be exploited for practical purposes to reduce it to a double-fold integral (on $b$ and $z$, if the wavefunction has spherical symmetry). In this case one gets
\begin{eqnarray}
G({\bf Q})&=&\int d^3 {\bf r} \ e^{-i{\bf Q.r}}\left<{\cal S}(b)\right>_Q
\psi_{jlm}({\bf r}) \nonumber \\
&=& 2\pi \int_0^\infty db\,  b \left<{\cal S}(b)\right>_Q 
\int_{-\infty}^\infty dz \exp[-iQ_z z]  {u_{lj}(r) \over r}\nonumber \\ &\times&\sum_{m_{l}\ m_{s}}(-i)^{m_l}{\cal C}_{lm_l}J_{m_l}(Q_t b)\left< lm_l sm_s | jm\right>\nonumber \\
&\times& P_{lm_{l}}\left(  \cos \vartheta\right)\chi_{m_s} ,\label{GQ}
\end{eqnarray}
where $Q_t$ and $Q_z$ are the transverse and longitudinal projections of ${\bf Q}$ along and perpendicular to the beam direction, $r=\sqrt{b^2+z^2}$ and $\cos \vartheta = z/\sqrt{b^2+z^2}$.
The inclusion of the spinors $\chi_{m_s}$ imply $|G({\bf Q})|^2\equiv G^\dagger G= |G_-|^2+|G_+|^2$, where $G_-$ is the spin-down and $G_+$ the spin-up amplitude. In the second equality we made use of the fact $Y_{lm} (\vartheta, 0) = C_{lm} P_{lm}(\cos \vartheta)$, where $P_{lm}$ are the Legendre polynomials, and
$$
{\cal C}_{lm}=(-1)^m \sqrt{2l+1\over 4\pi} \sqrt{(l-m)! \over (l+m)!}.
$$
$J_m(x)$ are cylindrical Bessel functions.

As a function of the total momentum ${\bf Q}$, another useful expression for $G({\bf Q})$ is
\begin{eqnarray}
G({\bf Q})&=&4\pi i^l \sum_{m_{l}\ m_{s}}i^{m_l}\left< lm_l sm_s | jm\right>Y_{lm_{l}}\left(  \widehat{\mathbf{Q}}\right)\chi_{m_s}\nonumber \\
&\times&\int dr \ r\ j_l(Qr)\left<{\cal S}(r)\right>_Q
 u_{lj}(r),\label{GQb}
\end{eqnarray}
where $j_l(x)$ is the spherical Bessel function.

\subsection{Cross sections}

Using the orthogonality relation of the spherical harmonics, we can integrate Eq.~\eqref{momdis} to get the {\it total momentum distribution}
 \begin{eqnarray}
{d\sigma\over Q^2d Q}& =& {2 \over \pi}   {S(lj)\over 2j+1}  \left< {d\sigma_{pN}\over d\Omega} \right>_{Q} \nonumber \\
&\times& \left| \int_0^\infty dr\,  r \left<{\cal S}(r)\right>_{Q_z} 
j_l(Qr) u_{lj}(r)  \right|^2 . \label{momtot}
\end{eqnarray}

\begin{figure}
[ptb]
\begin{center}
\includegraphics[
width=3.5in
]%
{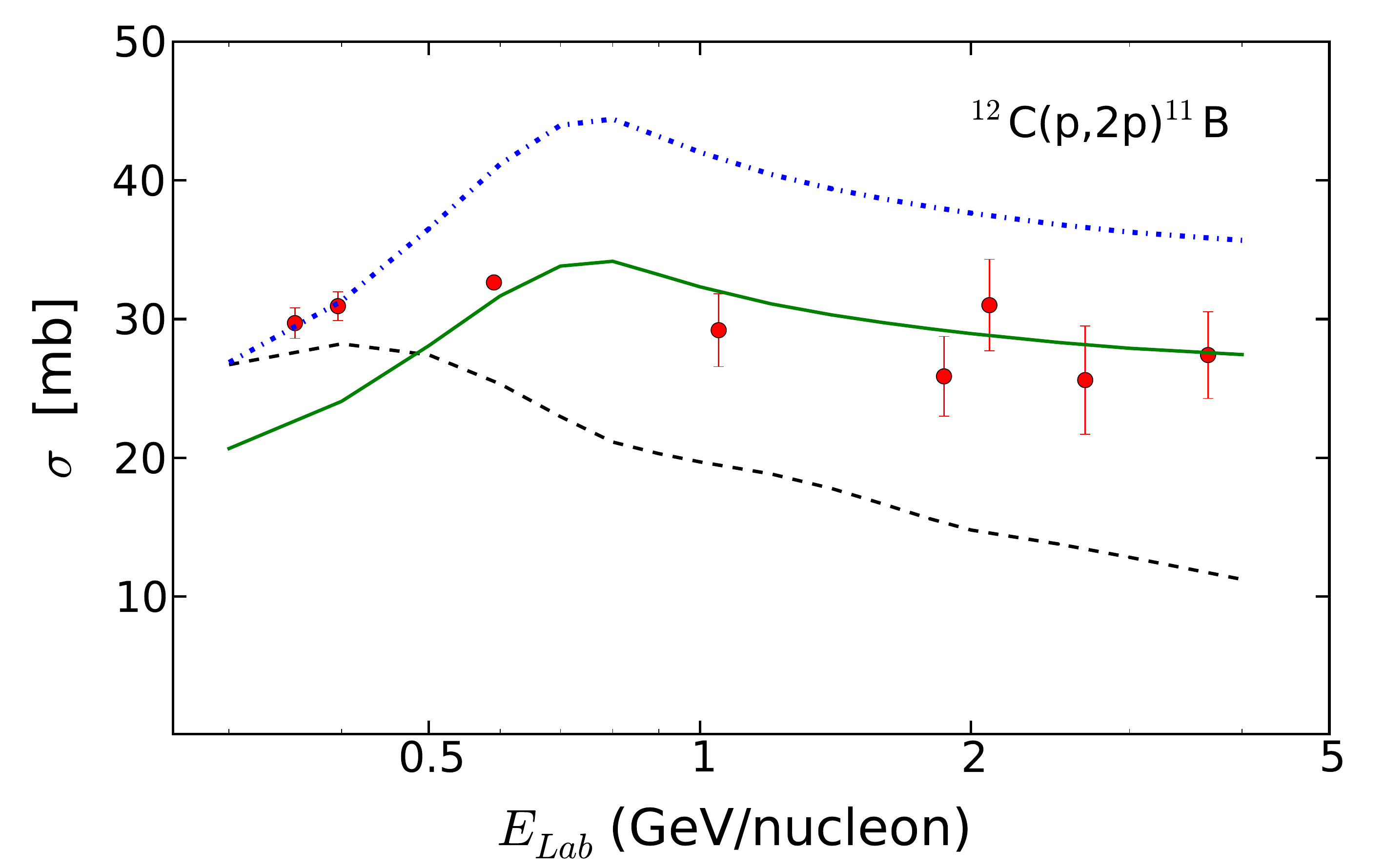}%
\caption{(Color online). Cross sections of $^{12}$C(p,2p)$^{11}$B fragmentation on hydrogen targets. The data are from Refs. \cite{Ols83,Web98}. The  dashed-dotted curve 
shows the cross section calculated according to Eq. \eqref{totxsec}, with  $\left< {d\sigma_{pp}/ d\Omega} \right>_{o.s.}=\sigma_{pp}^{tot}/4\pi$ where $\sigma_{pp}^{tot}$ is
total p-p free cross section.  The full curve shows the same calculation multiplied by a factor 0.77. The dashed curve uses  $\left< {d\sigma_{pp}/ d\Omega} \right>_{o.s.}$ given by the angular average of the {\it elastic} pp cross section satisfying the energy constraint of Eq. \eqref{cross2b}.}%
\label{12Cp2p}%
\end{center}
\end{figure}

Using the relation
\begin{equation}
\int_{0} ^{+ \infty} x J_m(\alpha x) J_m (\beta x) dx = {1\over \alpha} \delta(\alpha-\beta ), \label{deltt}
\end{equation}
we can integrate $|G({\bf Q})|^2$ over $Q_t$ using Eq. \eqref{GQ} and we obtain the {\it longitudinal momentum distribution},
 \begin{eqnarray}
{d\sigma\over d Q_z} &=&   {S(lj)\over 2j+1} \sum_m \left< {d\sigma_{pN}\over d\Omega} \right>_{Q_z} |{\cal C}_{lm}|^2 \nonumber \\
&\times& \int_0^\infty db\,  b \left| \left<{\cal S}(b)\right>_{Q_z} \right|^2 \nonumber \\
&\times&\left| \int_{-\infty}^\infty dz \exp[-iQ_z z]  {u_{lj}(r) \over r}P_{lm}(b,z)  \right|^2 . \label{momdisz}
\end{eqnarray}

Using the relation
\begin{equation}
{1\over 2\pi}\int \exp[i(\alpha-\beta)x] dx= \delta(\alpha-\beta ), \label{delzz}
\end{equation}
we can integrate $|G({\bf Q})|^2$ over $Q_z$ using Eq. \eqref{GQ} and we obtain the {\it transverse momentum distribution},
\begin{eqnarray}
&&{d \sigma \over Q_t d Q_t} =   {S(lj)\over 2j+1} \sum_m \left< {d\sigma_{pN}\over d\Omega} \right>_{Q_t} |{\cal C}_{lm}|^2 \nonumber \\
&\times& \int_{-\infty}^\infty dz \left| 
\int_0^\infty db  \left<{\cal S}(b)\right>_{Q_t}  {u_{lj}(r) \over r}J_{m}(Q_t b)P_{lm}(b,z)  \right|^2 . \nonumber\\
\label{momdist}
\end{eqnarray}

The {\it total cross section} is obtained either from integrations of Eq.~\eqref{momdisz} or of Eq.~\eqref{momdist}, using the closure relations \eqref{deltt} or \eqref{delzz}. One obtains
 \begin{eqnarray}
\sigma&=&   S(lj){2\pi \over 2j+1} \sum_m \left< {d\sigma_{pN}\over d\Omega} \right>_{o.s.} |{\cal C}_{lm}|^2 \nonumber\\
 &\times&\int_0^\infty db\,  b \left| \left<{\cal S}(b)\right>_{o.s.} \right|^2
 \int_{-\infty}^\infty dz  \left| {u_{lj}(r) \over r}P_{lm}(b,z)  \right|^2 . \nonumber \\
 \label{totxsec}
\end{eqnarray}
The subscript ``$o.s.$" in the above equation means that an average over the final momenta is made, which satisfies the on-shell conservation of energy for a non-relativistic  nuclear recoil energy, as described in Eq. \eqref{cross2b}.
In practice, the average is done by a Monte-Carlo sampling of the differential cross sections for several final momenta ${\bf k}_{p}^\prime$ and ${\bf k}_N$ with the constraint set by conservation of energy and momentum,
 \begin{eqnarray}
\left< {d\sigma_{pN}\over d\Omega} \right>_{o.s.} ={\displaystyle \int_{{\bf k}^{\prime}_{p},{\bf k}_N \in (o.s.)} d^3{\bf k}_p^{\prime} d^3{\bf k}_N  {\displaystyle d\sigma_{pN}\over d\Omega}  \over \displaystyle \int_{{\bf k}_{p}^{\prime},{\bf k}_N \in (o.s.)} d^3{\bf k}_{p} ^{\prime} d^3{\bf k}_N} \ .
\label{dselastdO}
\end{eqnarray}
For the momentum distributions obtained with Eqs.~(\ref{momtot},\ref{momdisz},\ref{momdist})  the average is done with the constraint set by the total momentum $Q$, while for the total cross section, the constraint $Q=0$ is used for simplicity.  

Eq. \eqref{totxsec} allows us to define a {\it nucleon knockout probability} at a given impact parameter $b$ by means of
 \begin{eqnarray}
P(b)&=&   {S(lj) \over 2j+1} \sum_m \left< {d\sigma_{pN}\over d\Omega} \right>_{o.s.} |{\cal C}_{lm}|^2 
 \left| \left<{\cal S}(b)\right>_{o.s.} \right|^2 \nonumber \\
&\times&  \int_{-\infty}^\infty dz  \left| {u_{lj}(r) \over r}P_{lm}(b,z)  \right|^2 . \nonumber \\
 \label{totxprob}
\end{eqnarray}
This equation is useful to calculate the parts of the transverse coordinate $b$ of the single-particle wavefunction which is probed by the nucleon
knockout mechanism.

The knockout of a nucleon will often lead to a state of short lifetime, which might decay to other channels than (p,pN). If, for example, an $0s_{1/2}$ proton from a nucleus like carbon has been knocked out, the Pauli principle no longer excludes those nuclear state rearrangements in which a  p-shell proton  falls into the hole of the s-shell ejecting another particle from the nucleus. This decay energy can also be shared among all nucleons forming a compound nuclear state which later decays by particle evaporation. For simplicity we will not consider these situations in this work, leaving it for future analysis.

\section{Results}
Before we discuss (p,2p) and (p,pn) reactions with unstable nuclei, we will establish how well the formalism described above reproduces quasi-free knockout reactions on stable targets. The ingredients for the calculations are the equations developed above plus (a) total nucleon-nucleon cross sections, using the fit of Ref.~\cite{BC10}, (b) elastic differential cross sections, using the Nijmegen global fit of the NN data base \cite{AW94}, (c) nuclear densities calculated with  the Hartree-Fock-Bogoliubov method, (d) single-particle wavefunctions calculated with a Woods-Saxon + spin orbit potential with parameters chosen to reproduce the separation energies, and (e) spectroscopic factors. According to Eq.~\eqref{totxsec}, total cross sections in (p,pN) reactions are directly related to spectroscopic factors. On the other hand, momentum distributions should depend on the angular momentum of the single-particle state probed in the reaction, according to Eqs.~(\ref{momtot},\ref{momdisz},\ref{momdist}). Unless stated otherwise, we use spectroscopic factors equal to the unity for the sake of concentrating on the dependence on other physical inputs in the calculations. 

\subsection{Stable nuclei}

\subsubsection{Cross sections}

In figure \ref{12Cp2p} we show the cross sections for $^{12}$C(p,2p)$^{11}$B fragmentation on hydrogen targets compared to experimental data.  
We consider p- (s)-states in $^{12}$C with proton separation energies of 15.9 (30.8) MeV and neutron separation energies of 18.7 (35.1) MeV, respectively. The dashed curve is a calculation using Eq. \eqref{totxsec} and  $\left< {d\sigma_{pp}/ d\Omega} \right>_{o.s.}$ given by the angular average of the {\it elastic} pp cross section satisfying the energy constraint of Eq. \eqref{cross2b}. For comparison purposes we also show by a dashed-dotted curve 
the cross sections calculated assuming  $\left< {d\sigma_{pp}/ d\Omega} \right>_{o.s.}=\sigma_{pp}^{tot}/4\pi$ where $\sigma_{pp}^{tot}$ is
total p-p free cross section.  In both cases the calculated cross sections fail to reproduce the experimental data. By using the total p-p cross section one also includes pion production which is likely to leave behind an excited fragment. One naturally expects that it will overestimate the value of the $^{12}$C(p,2p)$^{11}$B as compared to the experimental data. The full curve is the same as the dashed-dotted curve, but  multiplied by a factor 0.77. It shows a better agreement with the data. A rescaling by any factor would not  bring the dashed curve  to a good agreement with the experimental data.

Our calculations are done assuming the knockout from $p$-states only. Since the $s$-hole states are produced with high excitation energies, other particles
are expected to be emitted from the decay, leading to a small contribution to the (p,pN) channel \cite{Kob08}.

\begin{figure}
[ptb]
\begin{center}
\includegraphics[
width=3.5in
]%
{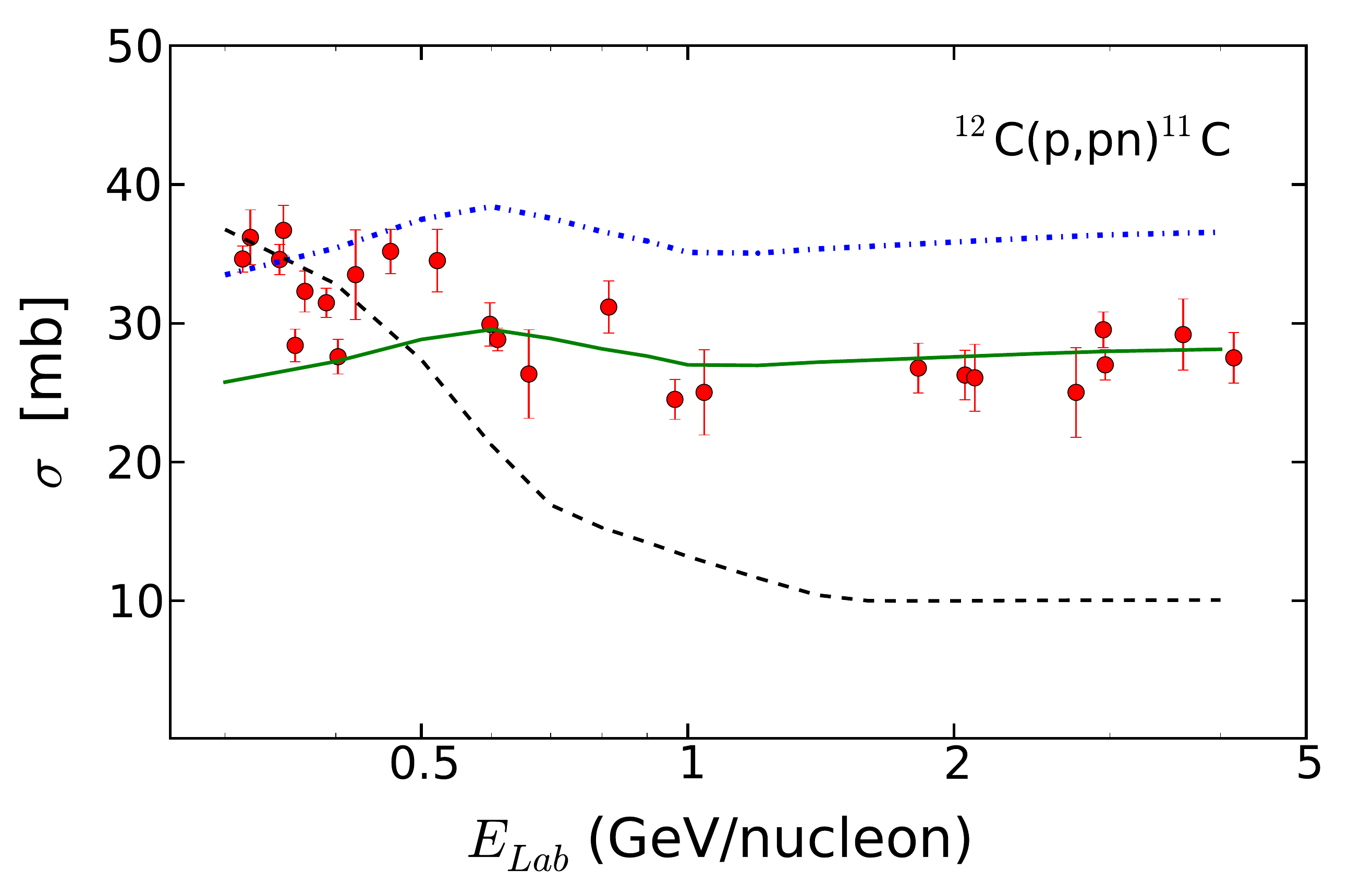}%
\caption{(Color online). Cross sections of $^{12}$C(p,pn)$^{11}$C fragmentation on hydrogen targets. The data are from Refs. \cite{Ols83,Web98,WK85,Web90,Cra56,Par60,Bur55,Ros56,Goe61,Cum58,HM60,Ben60,Kor02}. The  dashed-dotted curve 
shows the cross section calculated according to Eq. \eqref{totxsec}, with  $\left< {d\sigma_{pn}/ d\Omega} \right>_{o.s.}=\sigma_{pn}^{tot}/4\pi$ where $\sigma_{pn}^{tot}$ is
total p-n free cross section. The full curve shows the same calculation multiplied by a factor 0.77. The dashed curve uses  $\left< {d\sigma_{pn}/ d\Omega} \right>_{o.s.}$ given by the angular average of the {\it elastic} p-n cross section satisfying the energy constraint of Eq. \eqref{cross2b}.}%
\label{12Cppn}%
\end{center}
\end{figure}

In  figure \ref{12Cp2p} one observes a decrease of the calculated cross section as the bombarding energy decreases below $E_{lab} \lesssim 700$ MeV/nucleon. This is attributed to the fact that a sizable fraction of the scattered and knocked-out protons have an energy below 200 MeV. At these energies and below, the proton mean-free path is considerably reduced because the nucleon-nucleon cross sections rapidly increase as their relative energy  decreases. A model similar to ours, containing many simplifying assumptions, was published in Ref. \cite{Das72}. They have also used as input the total $\left< {d\sigma_{pp}/ d\Omega} \right>_{o.s.}=\sigma_{pp}^{tot}/4\pi$, with $\sigma_{pp}^{tot}$ being the p-p total cross section. Their results have a better agreement with other experimental data \cite{Das72} than ours. The reason for their successful results is puzzling as their model contains many more simplifying assumptions  than ours. 

The calculation using {\it elastic} p-p differential cross sections, with an average over the possible scattering angles, is shown by a dashed line in  figure \ref{12Cp2p}. It is very common in the literature to find calculations using $\left< {d\sigma_{pp}/ d\Omega} \right>_{o.s.}=\left( {d\sigma_{pp}/ d\Omega} \right)^{elast}_{\theta_{c.m.}=90^\circ}$. In our calculations we use the constraints set by the energy-momentum conservation laws, which allows for scattering angles close to but not restricted to this value. The (p,2p) cross sections calculated with elastic cross sections lie consistently below the experimental data for $E_{lab} \gtrsim 500$. And they deviate more from the experimental data with increasing bombarding energy. In fact, the elastic p-p cross section becomes increasingly smaller than $\sigma_{pp}^{tot}$ at energies beyond 500 MeV/nucleon \cite{PDG}. At these energies, the p-p angular distribution is also more and more anisotropic, showing a dip at c.m. scattering angles around 90$^\circ$. Hence, the  averaged p-p elastic cross sections include larger scattering angles, with outgoing protons being more strongly absorbed and with smaller scattering cross sections. 

\begin{figure}
[ptb]
\begin{center}
\includegraphics[
width=3.5in
]%
{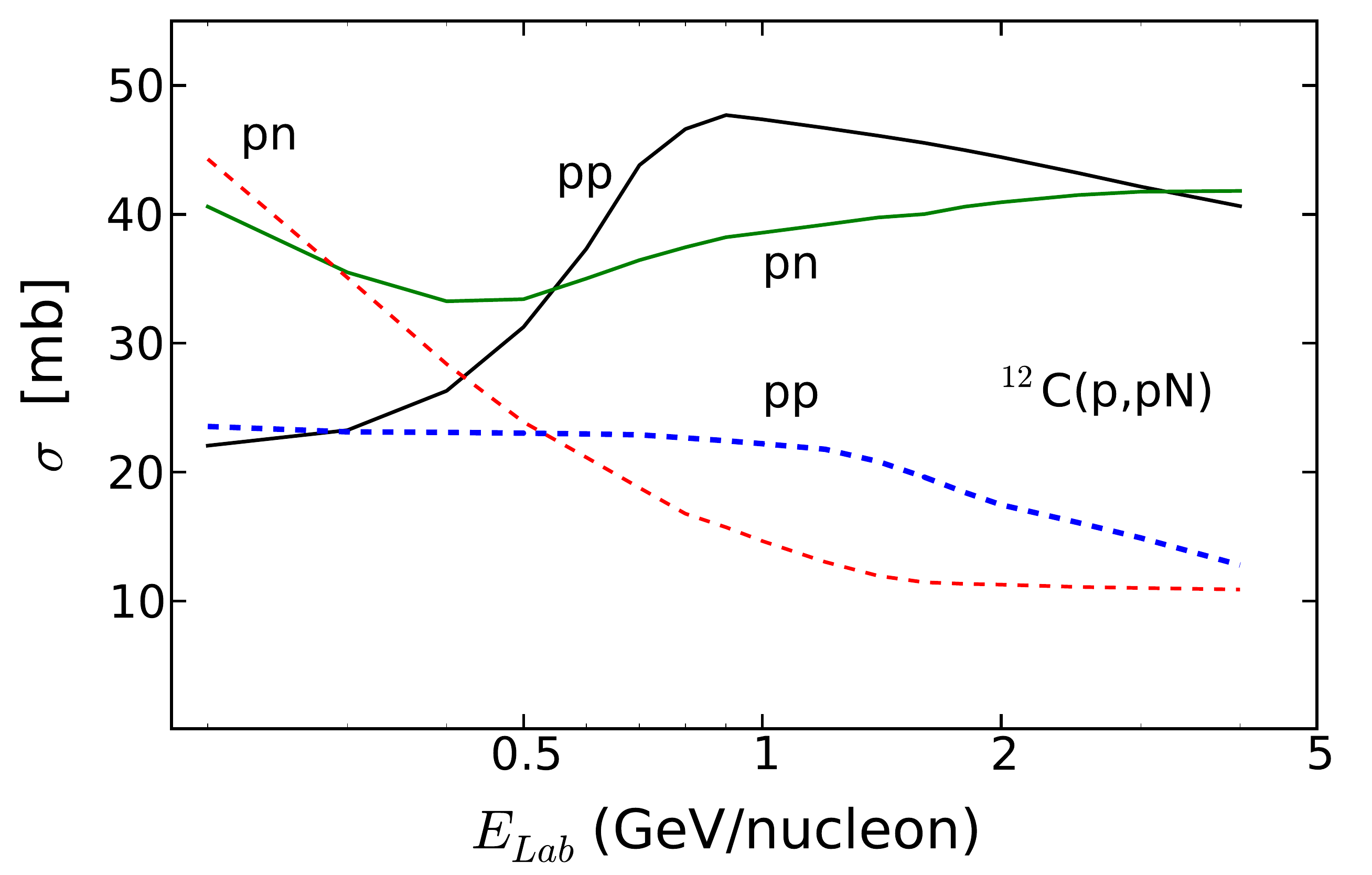}%
\caption{(Color online). Free p-p and p-n {\it total} cross sections (solid curves) as compared to the constrained angle averaged {\it elastic} p-p and p-n cross sections (dashed curves) according to Eq. \eqref{dselastdO} and for $^{12}$C(p,pN), with N = p or n. }%
\label{12CnnXs}%
\end{center}
\end{figure}

Similar features are seen our calculations presented in figure \ref{12Cppn} for $^{12}$C(p,pn)$^{11}$C as a function of the bombarding energy. Here the deviations obtained with the elastic p-n cross sections are more accentuated from those with isotropic {\it total} nucleon-nucleon cross sections than for the $^{12}$C(p,2p)$^{11}$B reaction (figure \ref{12Cp2p}). One obvious reason is that the p-n elastic angular distributions is more asymmetric than in the p-p case. At the lower energies, additional cross section from excitation of giant resonances (GRs) might contribute more, deserving further investigation.

In figure \ref{12CnnXs}  we show the free p-p and p-n {\it total} cross sections (solid curves) as compared to the constrained angle averaged {\it elastic} p-p and p-n cross sections (dashed curves) according to Eq. \eqref{dselastdO} and for $^{12}$C(p,pN), with N = p or n. Not all scattering angles are possible when the energy-momentum conditions in Eq. \eqref{cross2b} are met.  It is clear that the constrained angle averaged elastic cross sections are in most part responsible for the differences shown in figures \ref{12Cp2p} and \ref{12Cppn} between the knockout cross sections obtained with total and with elastic nucleon-nucleon cross sections. The other part responsible for the differences  is due to absorption effects generated by multiple binary collisions at different scattering angles. One observes in the figure that the medium averaged elastic cross sections (dashed curves) are smaller than the respective total free NN cross sections, as expected. However this difference decreases with decreasing energy and at energies below  300 MeV the elastic, medium averaged, cross sections become larger than the free total NN cross sections. This arises because absorption due to multiple scattering in the medium favors  scattering angles away from 90$^\circ$ degrees where the differential cross section has a minimum.   Hence, the averaged total elastic cross sections are larger than they would be without medium corrections because the largest values of the elastic differential cross section weight more on the average.

One of the main issues related to the (p,pN) cross sections at high energies ($\sim 1$ GeV) is to separate the different contributions arising from collective excitations leading to energy loss and to evaporation, and binary scattering with and without pion production.  As shown from direct experimental analysis (see, e.g., Ref. \cite{Web87}), the (p,pN) cross section gets contributions from (a) nucleon-nucleon quasi-elastic scattering, (b) nucleon-nucleon inelastic scattering, and (c) a low excitation energy. The cross sections, for all three cases are comparable. The shapes of the momentum distributions in cases can be reproduced with a nucleon-nucleon cascade model including pion production, but 
the theoretical predictions for the quasi-elastic component is too large by a factor of 3  \cite{Web87}. The low
momentum transfer peak is consistent with two mechanisms, (1) an excitation and decay via proton
emission of the carbon projectile and (2) a projectile proton scattering diffractively off the C target.
Finally, the Fermi momentum determined from the transverse momentum distribution is $160 \pm 11$
MeV/c compared to $190 \pm 11$ MeV/c from the longitudinal momentum distribution.

We thus conclude that a direct comparison of our calculations with experimental data as shown in
in Figures \ref{12Cp2p} and \ref{12Cppn} is hampered by the fact that the data are inclusive. A kinematically 
complete measurement of quasi-elastic scattering including the selection of the quasi-free kinematical condition
would allow a direct comparison with the calculated cross section using the elastic nucleon-nucleon cross
section as an input. However, such data have been measured so far only with very restricted angular acceptances. 
The analysis of Webb et al. \cite{Web87} shows that only around 50\% of the inclusive cross section at 1 GeV/nucleon
corresponds to quasi-elastic scattering. The cross section measured by those authors is 8.8(2.5) mb compared to our prediction of 
about 20 mb using elastic $d\sigma_{pp}/d\Omega$  at 1 GeV.  The reduction by 0.44(13) is to be compared to the usual reduction factor 0.6 from (e,e'p) experiments. 
The experimental cross section in Ref. \cite{Web87} has been acceptance corrected using the cascade model, meaning that it is model dependent.
The fully exclusive experiments in inverse kinematics as planned at the radioactive-beam facilities 
\cite{r3b,AU07,Kob08,PA12} 
will provide a full solid-angle coverage, and will thus be directly comparable to our predictions.

The features described in figures \ref{12Cp2p} and \ref{12Cppn} have also been observed for several other (p,2p) and (p,pn) in our studies. In the literature one finds experimental (p,pN) cross sections well described by means of phenomenological parameter fitting models, such as those presented in Refs. \cite{STB98,Web90}. The cross sections are also well described with microscopic theoretical models such as that reported in Ref. \cite{Das72} where total nucleon cross sections for Eq. \eqref{dselastdO} were used and an analytical formulation for the (p,pN) reaction was given, with further approximations.  We are not certain what physics input in such models validate their use and allow for a better agreement with the data. 

Other common approximations found in the literature are (a) replacing $\left< {d\sigma_{pN}/ d\Omega} \right>_{o.s.} $ by the {\it total} free nucleon-nucleon cross section at 90$^\circ$ and keeping the same value for all bombarding energies (see, e.g., Ref. \cite{Ma58,{JM73,MTV79}}), and (b) using phenomenological effective nucleon-nucleon interactions fitted to other nucleus-nucleus collision processes (such as charge-exchange reactions), instead of nucleon-nucleon cross sections (see, e.g., Ref. \cite{KM86}). The use of Glauber theory was also recognized as an important theoretical tool to treat multiple scattering as well as to formulate a relativistic covariant model (see, e.g., Refs. \cite{Ove06,Ove07,CR09,RCV11}). On the other hand, the use of Mandelstam variables has proven to have the advantage  of relating theoretical values to detection efficiencies in high-energy collisions \cite{Ste80} in a more straightforward way. This approach has been used in Ref. \cite{Chu05} to study quasi-free $\alpha$ knockout from $^{6,8}$He beams with a relative success.

It is also evident from figures \ref{12Cp2p} and \ref{12Cppn} that the experimental data at energies of the order of 200-500 MeV are rather well described by using either total or elastic nucleon-nucleon cross sections. In fact, the total elastic and total inelastic nucleon-nucleon cross sections in free space have nearly the same values at this energy range. The differential elastic cross sections tend to be more asymmetric at larger energies. We therefore conclude that this energy range should be best suitable for studies of (p,pN) reactions.

\begin{figure}
[ptb]
\begin{center}
\includegraphics[
width=3.5in
]%
{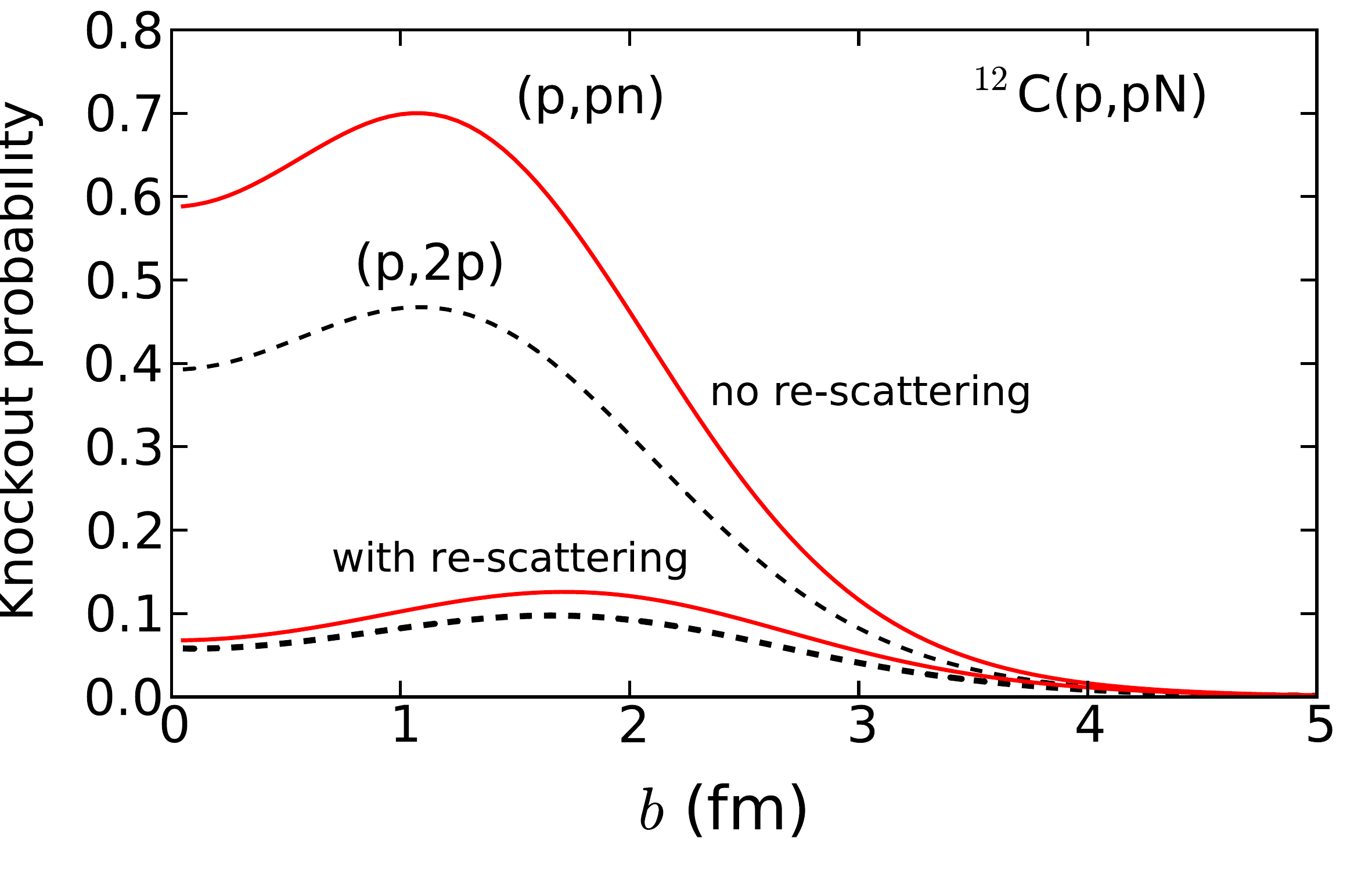}%
\caption{(Color online). Proton and neutron knockout probabilities from p-states in $^{12}$C at 500 MeV/nucleon, according to Eq. \eqref{totxprob}. The solid curves are for (p,pn) and dashed curves for (p,2p). We also show the results obtained with $\left<{\cal S}(b)\right>_{o.s.} =1$ in Eq.  \eqref{totxprob} (no re-scattering).  }%
\label{probf}%
\end{center}
\end{figure}

In figure \ref{probf} we plot the proton and neutron knockout probabilities from p-states in $^{12}$C using Eq. \eqref{totxprob}. The solid curves are for (p,pn) and dashed curves for (p,2p). We also show the results obtained with $\left<{\cal S}(b)\right>_{o.s.} =1$ in Eq.  \eqref{totxprob}. This would correspond to a direct NN knockout, without multiple binary collisions (no re-scattering). In this case the knockout probabilities are very large, almost violating unitarity. Absorption due to knockout reduces the one-nucleon knockout probability considerably, as can be seen by the lower curves. There is a larger reduction at smaller impact parameters than at the surface, as expected from basic principles. From central collisions ($b=0$) there is a larger reduction  due to absorption from the one-nucleon knockout channel. For collisions at low impact parameters we find an average reduction by a factor 6.8 for (p,2p) reactions and by a factor 8.7 for (p,pn) reactions. 

\begin{figure}
[ptb]
\begin{center}
\includegraphics[
width=3.3in
]%
{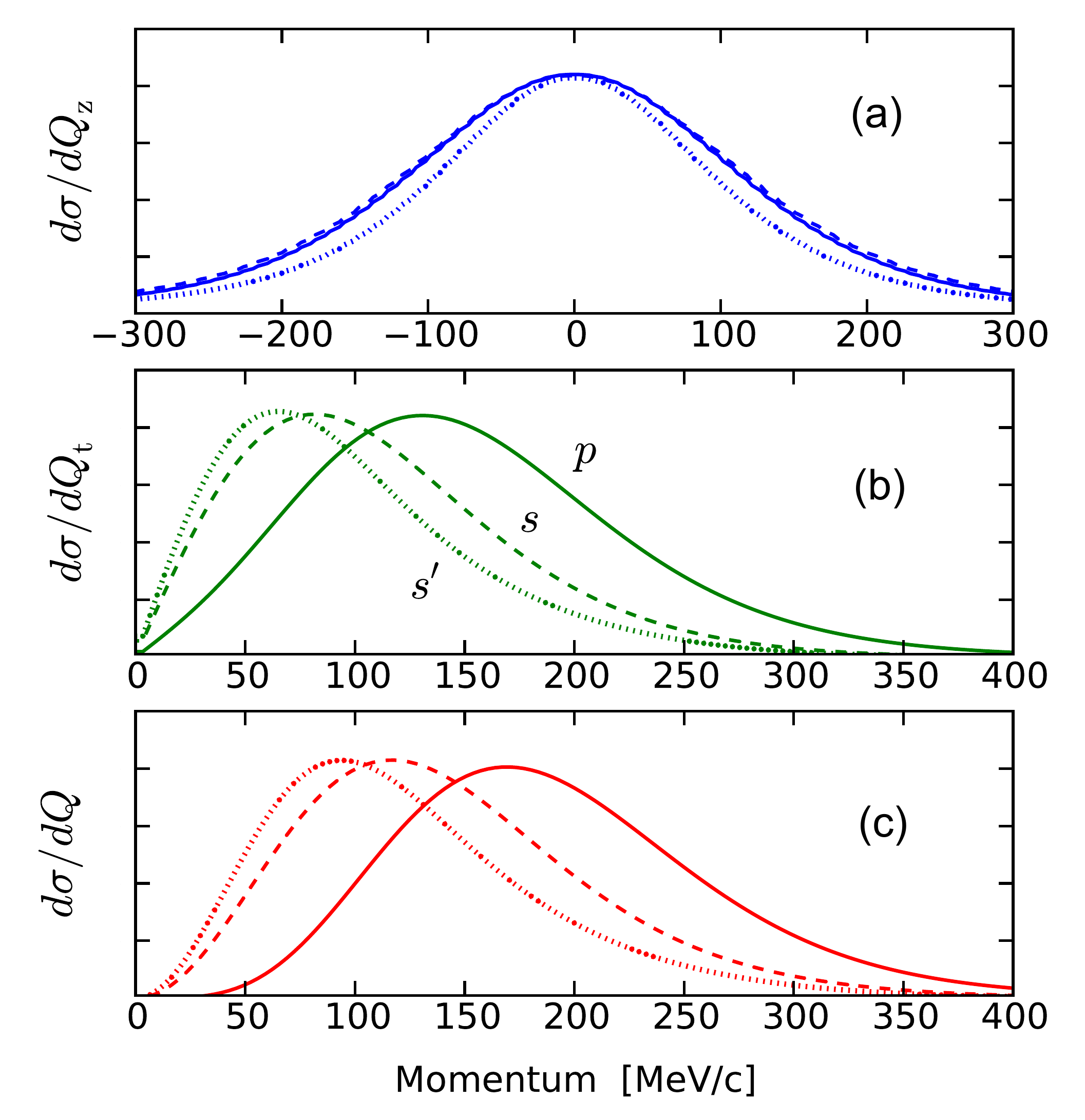}%
\caption{(Color online). Longitudinal (upper panel), transverse (middle panel) and total momentum distributions (lower panel) for the missing momentum in $^{12}$C(p,2p)$^{11}$B fragmentation on hydrogen targets. The  full curves are for $p$-states, the dashed curves are for $s$-states. The dotted curves $s^\prime$ correspond to the calculations where the energy of the $s$-states are taken as the  $p$-state energy. The curves are rescaled to be shown in the same plot.}%
\label{12Cldep}%
\end{center}
\end{figure}

Multiple scattering effects have been studied by several authors, using the Feshbach-Koonin-Kawai formalism (see, e.q., \cite{Cow80,Cia83,Cia84,Cow88,Pil89,Cow90,For93,Cia84,Cow98,Stey99,FKK80}), or the multiple scattering Glauber formalism (see, e.g., Refs. \cite{Ove06,Ove07,CR09,RCV11}), or even with a simple one parameter rescaling factor model (see, e.g. Refs. \cite{JM73,Das72}). It is clear however that the effects of multiple scattering are very large and need to be taken into account carefully if the total cross section is supposed to be useful for the purposes of extracting spectroscopic factors.  

A simple approximation for dealing with the effects of multiple scattering is to assume that the central probability scales as 
\begin{equation}P\simeq P_0 \exp \left[ - {3R\over \lambda} \right]\end{equation}
where $P_0$ is the probability without absorption, $R$ is the nuclear radius,  and $\lambda$ is the nucleon mean free path. The factor 3 in the numerator is due to the three particles; one in the incoming channel and two in the outgoing channel.  We use $R=2.6$ fm, corresponding to the mean square radius of $^{12}$C. This yields a mean free path for protons in $^{12}$C of about $\lambda \simeq 2.9$ fm at 800 MeV/nucleon. Despite the approximations used for this estimate, this result is remarkably close to the mean-free-path of high-energy  protons in $^{12}$C, $\lambda \simeq 2.4$ fm, deduced in Ref.~\cite{Tan81}.  

\subsubsection{Momentum distributions}

In nucleon knockout in heavy-ion reactions, momentum distributions are known to be a useful probe of the angular momentum of the knocked-out nucleon. Due to the centrifugal barrier, a state with large $l$ is confined to a smaller region, yielding a broader momentum distribution \cite{HBE96}. The separation energy is also an important factor to determine the size of the bound-state single-particle function and has been a useful tool to uncover the existence of unstable halo nuclei \cite{Tan85}.  In our calculations, the proton $p$-state energy in $^{12}$C is set to the proton separation energy, i.e., $\epsilon_p= - 15.9$ MeV and the $s$-state energy to $\epsilon_s = -30.8$ MeV. 

In figure \ref{12Cldep} we show the longitudinal (upper panel), transverse (middle panel) and total momentum distributions (lower panel) for the missing momentum, Eq. \ref{cross2b}, in $^{12}$C(p,2p)$^{11}$B reactions on hydrogen targets at 500 MeV/nucleon. The  full curves are for $p$-states, the dashed curves are for $s$-states. The dotted curves $s^\prime$ correspond to calculations for which the energy of the $s$-states are taken as the same energy as the $p$-state energy. We have normalized the curves to a peak value as we are only interested in the variation of the  widths of the distributions with the energy and angular momentum of the involved single-particle orbitals.

Two effects are competing in the calculations of the $p$-state and $s$-state proton removal shown by the solid and dashed curves  in figure \ref{12Cldep}. On the one hand, the $l$-dependence widens the distribution for the $p$-states compared to those for the $s$-states. On the other hand, the separation energy of the $s$-state being larger than that for the $p$-state compensates by widening the $s$-state contribution. The transverse and total momentum distributions (middle and low panel) are still able to discern between the two states. But for the longitudinal momentum distributions the two curves are almost identical in shape.  When the energy of the $s$-state is artificially set equal to the energy of the $p$-state, the shape of the $s$ and $p$ momentum distributions differ substantially. It is clear that the momentum distributions displayed either in terms of  transverse, or total missing momentum, are a sensitive probe of the angular momentum state of the orbital from which the proton is removed. A similar conclusion was obtained in (e,e$^{\prime}$p) momentum distribution analyses \cite{Mou76}. 

\subsection{Neutron rich nuclei}

The prominent differences between stable and neutron-rich unstable nuclei are: (a) extended neutron distribution in the form of a halo, or skin, and (b) smaller binding energies than those for stable nuclei. Here we will explore both cases and their imprints in the cross sections and momentum distributions in (p,2p) and (p,pn) reactions.

\subsubsection{Cross sections}

As an example for the dependence of the cross sections on the separation energy, we consider (p,pn) reactions with $^{23}$O, with  valence neutrons in the $[0d_{5/2}]^6$ $[1s_{1/2}]$ configuration. The separation energies are   $S_n(1s_{1/2}) = 2.73$ MeV and $S_n(0d_{5/2}) = 6.0$ MeV. We artificially vary the separation energy of theses states to explore the separation energy dependence of the (p,pn) cross sections at 500 MeV/nucleon.

\begin{figure}
[ptb]
\begin{center}
\includegraphics[
width=3.5in
]%
{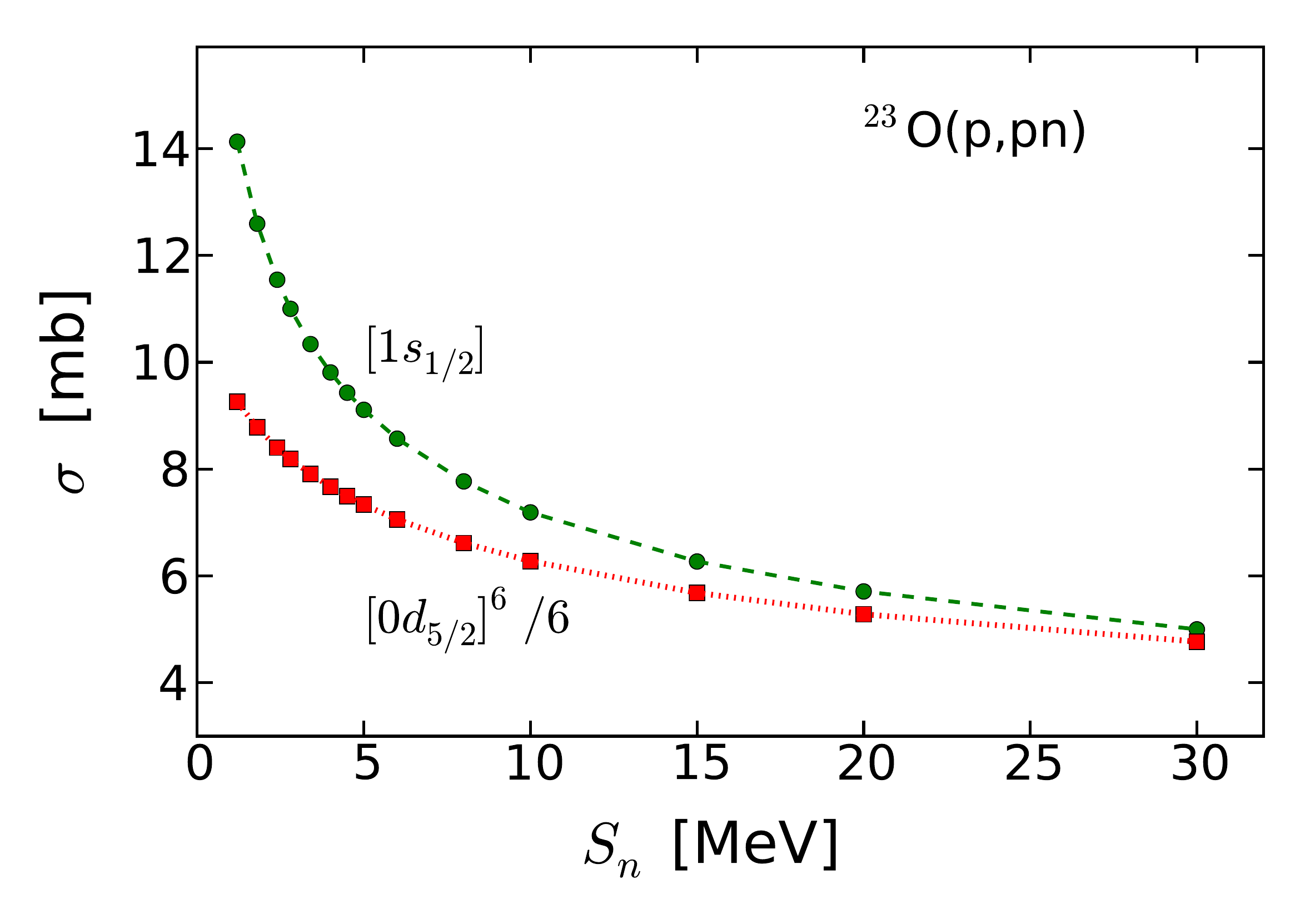}%
\caption{(Color online). Cross sections for neutron removal in (p,pn) reactions on $^{23}$O from $[0d_{5/2}]^6$ and $[1s_{1/2}]$ orbitals as a function of the separation energy. The cross sections for neutron removal from the $[0d_{5/2}]^6$ orbital is divided by the number of the neutrons in the orbital (6). The separation energies are varied artificially. The dashed and dotted curves are guide to the eyes.}%
\label{o23XS}%
\end{center}
\end{figure}

In figure \ref{o23XS} we plot the cross sections for neutron removal in (p,pn) reactions on $^{23}$O from $[0d_{5/2}]^6$ and $[1s_{1/2}]$ orbitals as a function of the separation energy. The cross sections for neutron removal from the $[0d_{5/2}]^6$ orbital is divided by the number of the neutrons in the orbital (6). As expected, the cross sections are strongly energy dependent close to the threshold and steadily decrease with increasing separation energy. Close to threshold (i.e., close to $S_n=0$) a large chunk of the wavefunction lies in a region where absorption, or multiple scattering, is smaller, thus increasing the removal probability at larger impact parameters, consequently increasing the knockout cross section. As the separation energy increases it becomes harder to knockout a neutron without rescattering effects.

\begin{figure}
[ptb]
\begin{center}
\includegraphics[
width=3.5in
]%
{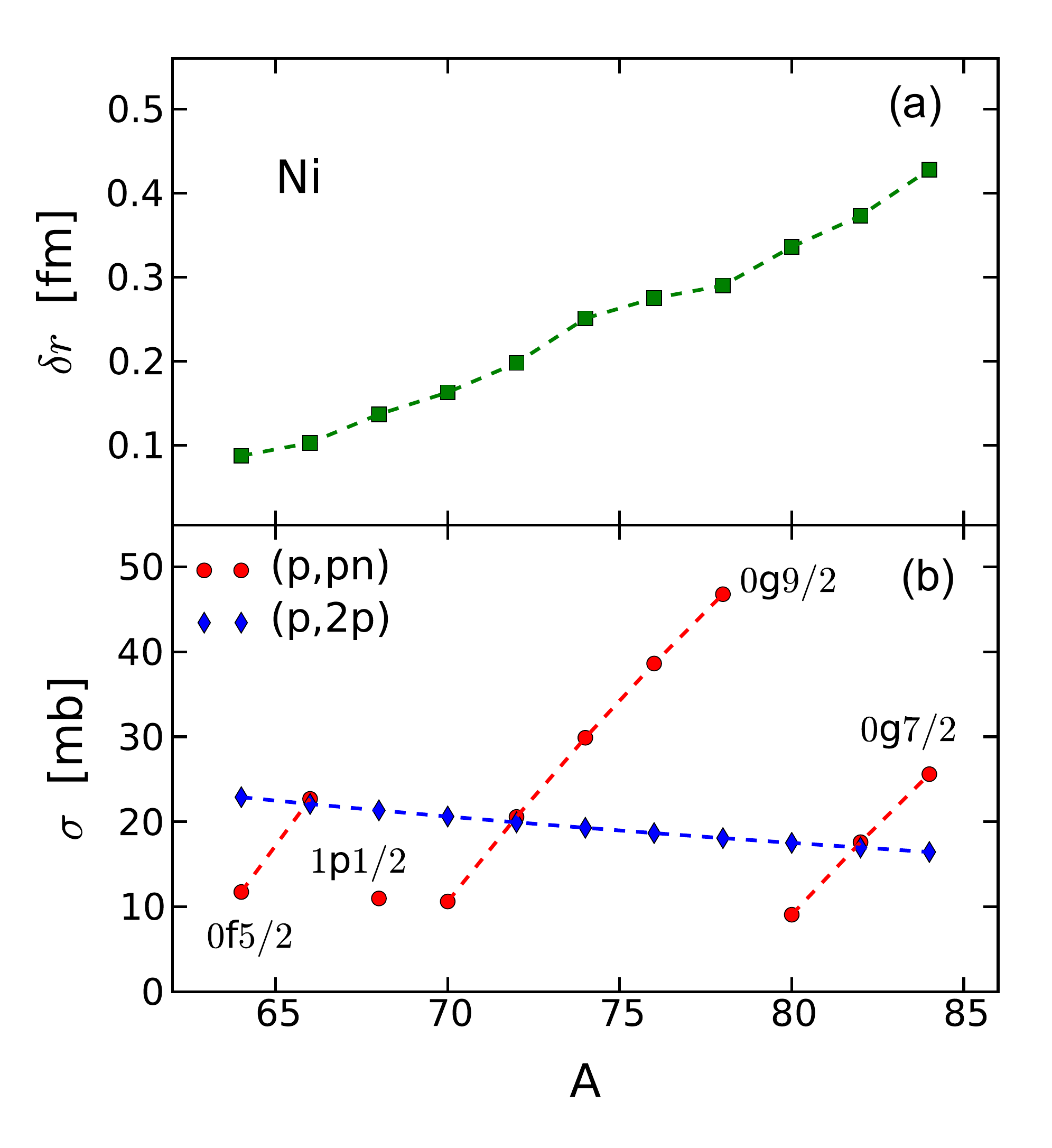}%
\caption{(Color online). Neutron skin of even-even Ni isotopes obtained with a HFB calculation (upper panel). The lower panel displays the (p,pn) and (p,2p)  cross sections on Ni isotopes at 500 MeV/nucleon assuming a fixed binding energy for the knocked out nucleon (see text for explanation).  The dashed curves are guide to the eyes.}%
\label{NiSkin}%
\end{center}
\end{figure}

We play a similar game as above in order to test the dependence of the cross sections on the matter distribution. To avoid inclusion of the dependence of the cross section on the separation energy, we assume a constant separation energy for the removed neutrons and protons. We also adopt the naive shell-model with 8 protons in the $0f_{1/2}$ shell.  The neutrons fill progressively the $0f_{5/2}$, $1p_{1/2}$, $0g_{9/2}$ and $0g_{7/2}$ orbitals.  The neutron and proton densities for the Ni isotopes are calculated within the Hartree-Fock-Bologliubov model using the SLy4 interaction \cite{Cha98} and a mixed pairing interaction (see Ref. \cite{Ber09}). Only even-even isotopes were considered. The neutron skin thickness is characterized by the difference of neutron and proton rms radii,
\begin{equation}
\delta r = \left< r_n^{2} \right>^{1/2}-\left< r_p^{2} \right>^{1/2} .
\end{equation}
In figure \ref{NiSkin} we show our results for the neutron skin of even-even Ni isotopes in the upper panel. The lower panel displays the (p,pn) and (p,2p)  cross sections obtained assuming a fixed binding energy for the knocked out nucleon.

It is clear from figure \ref{NiSkin} that the increasing neutron skin has a small effect on the (p,2p) cross sections, if the separation energy of the removed proton is kept constant. The main effect is to reduce the cross section due to the larger rescattering probability as the nuclear mass increases. The neutron-proton and proton-proton total cross sections are about the same at this energy.  The large nuclear charge  and the Coulomb barrier for the proton leads to a concentration of the proton wave function at the nuclear center where the absorption effects are stronger, yielding to a larger reduction of the (p,2p) cross section compared to (p,pn). However, the increasing  neutron numbers in each orbital will increase the neutron removal probabilities as the orbital is filling up. This is clearly seen in the figure.  In principle, the cross sections for neutron removal would increase even faster because an increasing neutron skin often means a smaller separation energy for a valence neutron.   As we discussed in connection with figure \ref{o23XS}, decreasing separation energies lead to increasing values of removal cross sections. Therefore, the dependence on the neutron skin should be manifest throughout the increasing number of neutrons in the orbitals and also with the separation energy of the removed neutron. For proton removal the increasing neutron skin has a lesser important role on the separation energy.  

It is worthwhile to compare which parts of the wave function are accessible in knockout reactions with heavy ions with those obtained in (p,pN) reactions.  The theory for knockout reactions with heavy ions, routinely used to extract nuclear spectroscopic information, relies on the  eikonal theory developed in Ref.~\cite{HM85}. The theory is based on  probability arguments using the eikonal S-matrix to obtain the parts of the nucleon wave function which are ``measured" by the reaction mechanism. It has later been proven to be a valuable tool for reactions involving unstable nuclear beams \cite{BM92,HBE96,GT10}.   Following the same formalism as in Refs. \cite{HM85,BM92,HBE96} one obtains for the probability to remove a nucleon in orbital ($jl$) located at distance ${\bf b}$ (perpendicular to the collision axis) from the center of the projectile as
\begin{eqnarray}
 &&P_{jl}(b) = {1 \over 2l+1} \left| S_c(b)\right|^2 \sum_m  \int d^3 {\bf r} \left| \psi_{jlm}({\bf r})\right|^2  \nonumber \\
 &\times& \left[ 1- \left|S_n \left( \sqrt{r^2 \sin^2 \theta + b^2 -2rb \sin \theta \cos \phi}\right)\right|^2 \right] . \nonumber \\
 \label{pjl}
 \end{eqnarray}
 Here $S_N$ ($S_c$) is the eikonal matrix amplitude for the scattering  of the nucleon N (core c) on the target, and ${\bf r} \equiv (r, \theta, \phi)$.

\begin{figure}
[t]
\begin{center}
\includegraphics[
width=3.3in
]%
{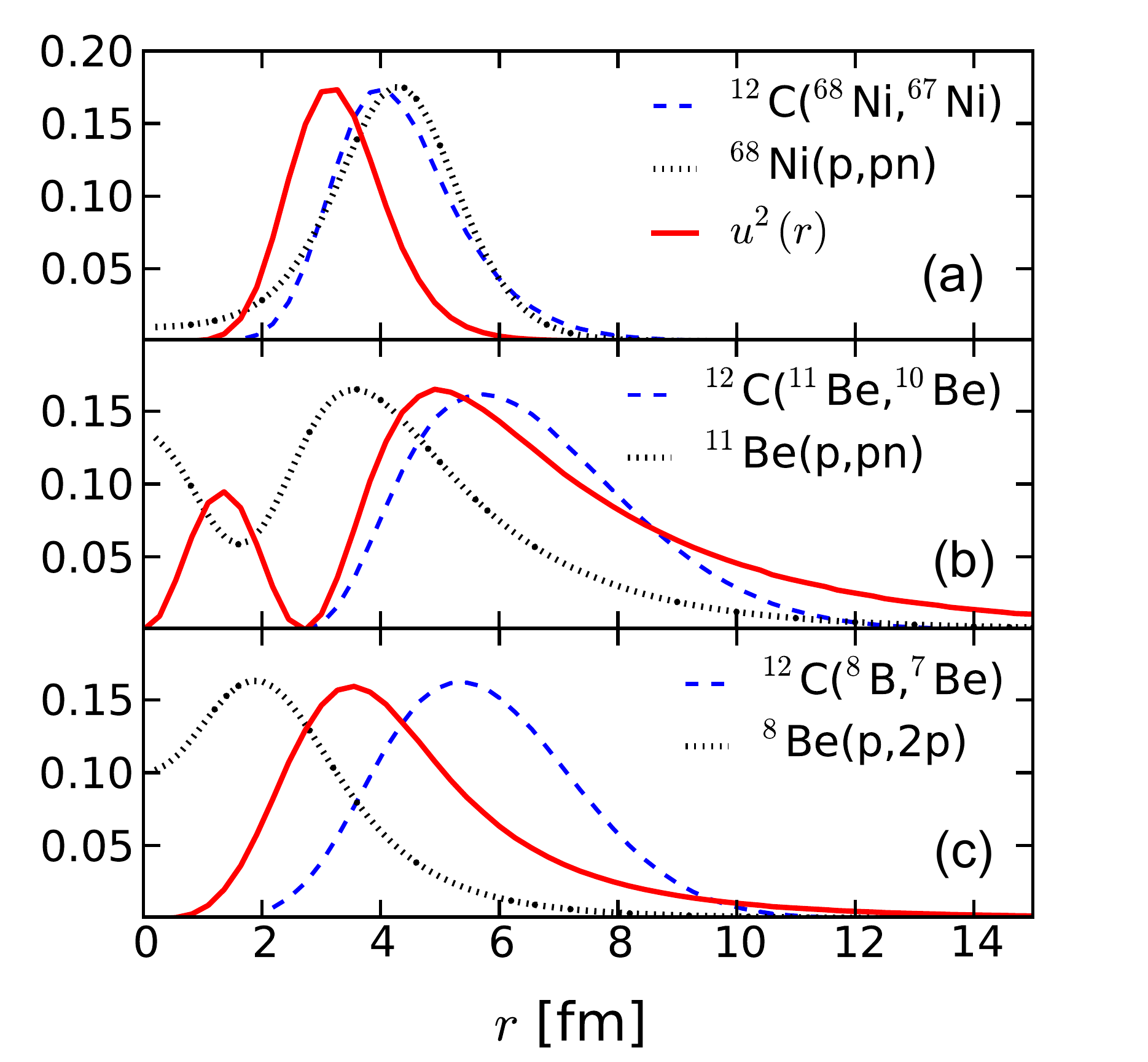}%
\caption{(Color online). {\it Upper panel:} The dashed curve shows the probability for removal of a neutron in the reaction $^{12}$C($^{68}$Ni,$^{67}$Ni) at 500 MeV/nucleon as a function of the distance to the c.m. of $^{68}$Ni. The dotted curve represents the removal probability in a $^{68}$Ni(p,pn) reaction at the same energy. For comparison the square of the radial wave function $u(r)$ is also shown (solid curve). {\it Middle panel:}  Same as the upper panel, but for the reactions $^{12}$C($^{11}$Be,$^{10}$Be)  and $^{11}$Be(p,pn) at 500 MeV/nucleon. {\it Lower Panel:}  Same as upper panel, but for the reactions $^{12}$C($^{8}$B,$^{7}$Be)  and $^{8}$B(p,2p) at 500 MeV/nucleon. }%
\label{probabilities}%
\end{center}
\end{figure}

In figure \ref{probabilities}, upper panel, we show the dashed curve shows the probability for removal of a neutron in the reaction $^{12}$C($^{68}$Ni,$^{67}$Ni) at 500 MeV/nucleon as a function of the distance to the c.m. of $^{68}$Ni. The dotted curve represents the removal probability in a $^{68}$Ni(p,n) reaction at the same energy. For comparison the square of the radial wave function $u(r)$ is also shown (solid curve). We assume a neutron in the $0f_{7/2}$ orbital in $^{68}$Ni, bound by 15.68 MeV. The figures in the middle panel are for the reactions $^{12}$C($^{11}$Be,$^{10}$Be)  and $^{11}$Be(p,n) at 500 MeV/nucleon.   We assume a neutron in the $1s_{1/2}$ orbital in $^{11}$Be, bound by 0.54 MeV. The figures in the lower panel are for the reactions $^{12}$C($^{8}$B,$^{7}$Be)  and $^{8}$Be(p,2p) at 500 MeV/nucleon.   We assume a neutron in the $0p_{3/2}$ orbital in $^{8}$Be, bound by 0.14 MeV. One observes that the removal cross sections for both knockout and (p,pn) reactions probe the surface part of the wave function. This is due to the fact that in  both cases, the absorption is very strong for small impact parameters. For proton or neutron removal from even deeper bound states, with a concentration of the wave function closer to the origin,  both reaction mechanisms will probe an even smaller part of the wave function tail.

In the middle panel of figure \ref{probabilities} one sees that  the part of the wave function probed in the $^{12}$C($^{11}$Be,$^{10}$Be) is again limited to the surface of the nucleus, beyond the orbital maximum density. On the other hand, the (p,pn) reaction has a much larger probability of  accessing information on the inner part of the wave function, as seen by the dashed curve. 
These results are in agreement with the conclusions drawn in Ref. \cite{RCV11} for stable nuclei where it has been shown that for light nuclei the average density probed in (e,e'p) is comparable to the one probed in (p,2p). There is a strong A dependence, though, and for medium-heavy and heavy nuclei one is rather probing the surface region in (p,pN) reactions.
It is thus clear that knockout reactions with heavy ions and (p,pN) reactions yield complementary nuclear spectroscopic information. For deep bound states the first reaction is only accessible to the tail of the nuclear wave function, whereas the (p,pN) reaction process probes the largest part of the  wave function for loosely bound nuclei.

\subsubsection{Momentum distributions}

\begin{figure}
[t]
\begin{center}
\includegraphics[
width=3.5in
]%
{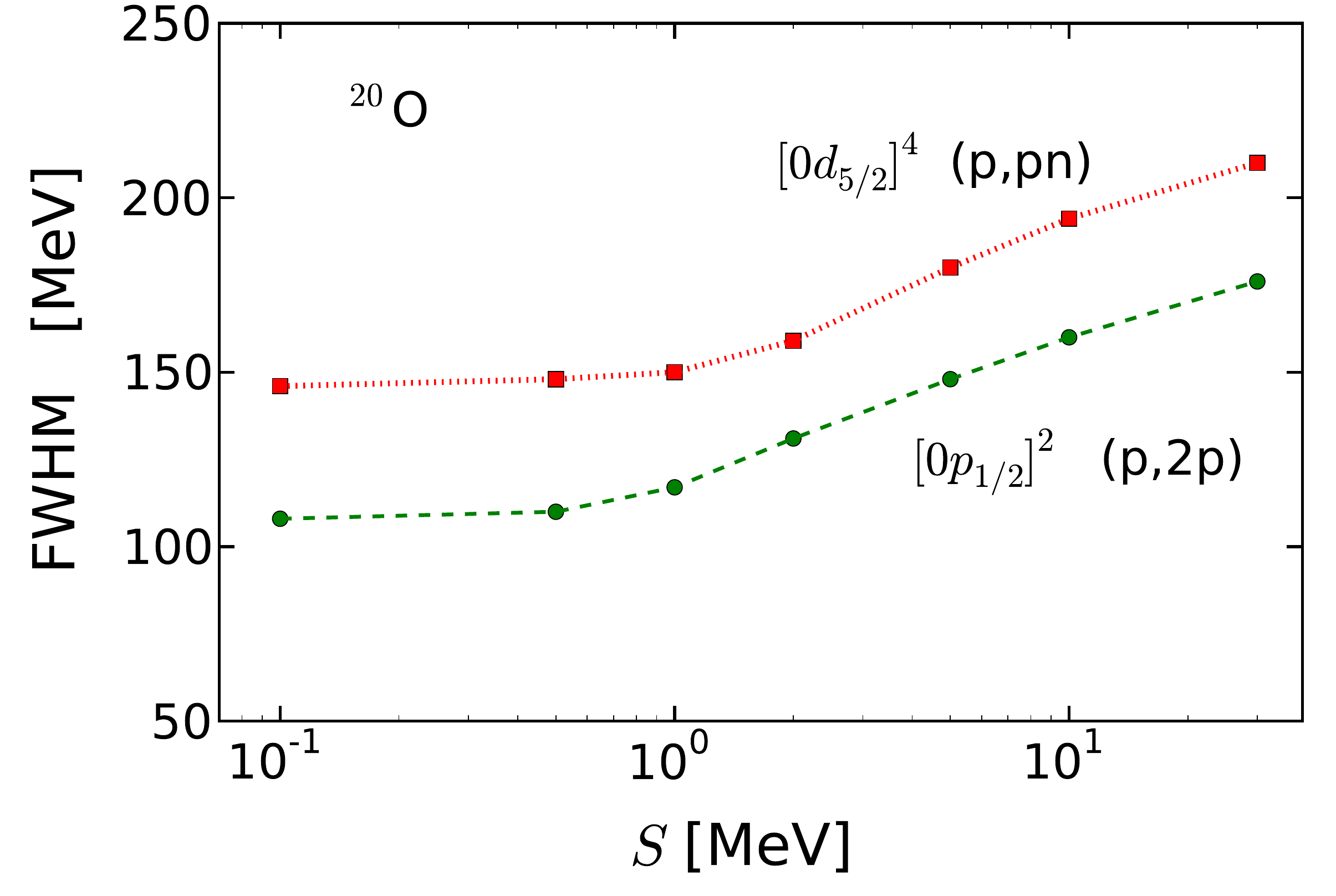}%
\caption{(Color online). Full width at half maximum (FWHM) of the transverse momentum distributions in (p,2p) and (p,pn) reactions of $^{20}$O at 500 MeV/nucleon as a function of the proton and neutron separation energies. The upper curve is for (p,pn) and lower curve for (p,2p) reactions. The separation energies are varied artificially (see text for explanation). The dashed and dotted curves are guide to the eyes.}%
\label{width}%
\end{center}
\end{figure}

The shape of the momentum distributions for (p,2p) and (p,pn) reactions always have similar characteristics as those shown in figure \ref{12Cldep}, independent of the nucleus. As we have discussed before, for stable nuclei the shape will depend mainly on the angular momentum and on the separation energy of the knocked out nucleon. The dependence on the angular momentum is obvious, following the same trend for either stable or unstable projectiles. But unstable projectiles often exhibit very low nucleon separation energies. As with nucleon removal reactions with heavy ions, one also expects that the width of the momentum distribution are strongly dependent on the separation energy,  in particular close to the dripline. We will test this with $^{20}$O which has a proton (neutron) separation energy of 19.43 (7.61) MeV. We will artificially vary these values to learn how the widths of the transverse momentum distributions will vary with the separation energy. The knocked out protons and neutrons are assumed to occupy the $[0p_{1/2}]^2$  and $[0d_{5/2}]^4$ levels, respectively. 

In figure \ref{width} we show the full width at half maximum (FWHM) of the transverse momentum distributions in (p,2p) and (p,pn) reactions of $^{20}$O at 500 MeV/nucleon as a function of the proton and neutron separation energies. The upper curve is for (p,pn) and lower curve for (p,2p) reactions. As expected, the widths for d-states are larger than those for p-states.

\begin{figure}
[t]
\begin{center}
\includegraphics[
width=3.in
]%
{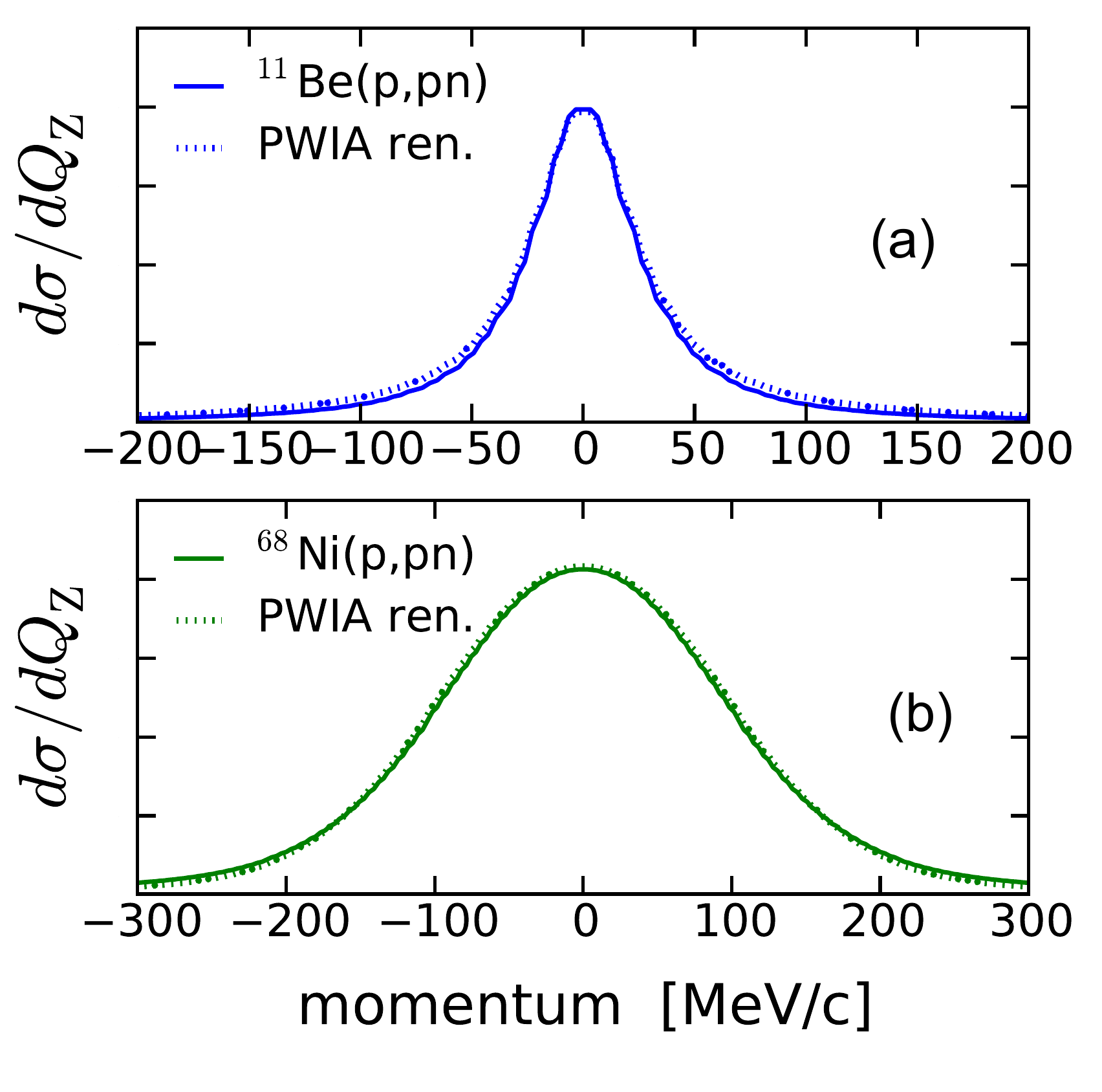}%
\caption{(Color online). Longitudinal  momentum distributions for $^{11}$Be(p,n) and and $^{68}$Ni(p,n) reactions at 500 MeV/nucleon. For $^{11}$Be  we assume neutron removal from the 1s$_{1/2}$ states with 0.502 MeV separation energy, while for $^{68}$Ni we assume the removal from 0f$_{3/2}$ orbital with separation energy of 15.68 MeV. PWIA results are shown as dashed lines, whereas the solid lines are for the full DWIA.}%
\label{widthqz}%
\end{center}
\end{figure}

A very different trend exists for (p,pN) reactions in comparison to nucleon knockout reactions induced by heavy ions. For the latter case, the width of the momentum distributions decreases strongly with decreasing separation energies. That was in fact, one of the hallmarks for the identification of the first halo nuclei \cite{Tan85}. In contrast, one observes in figure \ref{width} a ``saturation" of the width for decreasing separation energy. The reason for this behavior is the fact that knockout reactions induced by heavy ions are a very peripheral process, being almost entirely sensitive to the tail of the nucleon wave functions \cite{BM92}.  In contrast, (p,pN) reactions are in great part influenced by the part of the nucleon wavefunction inside the nucleus. The width of the momentum distributions for (p,pN) reactions reduce with decreasing separation energies because an extended part of the nucleon wave function is accessed as the separation energy decreases till a ``saturation" of the width is reached at low separation energies. As concluded from our calculations (see, e.g., figure \ref{o23XS}), the total cross section is a better measure of the nucleon separation energy in loosely bound nuclei. 

\begin{figure}
[t]
\begin{center}
\includegraphics[
width=3.in
]%
{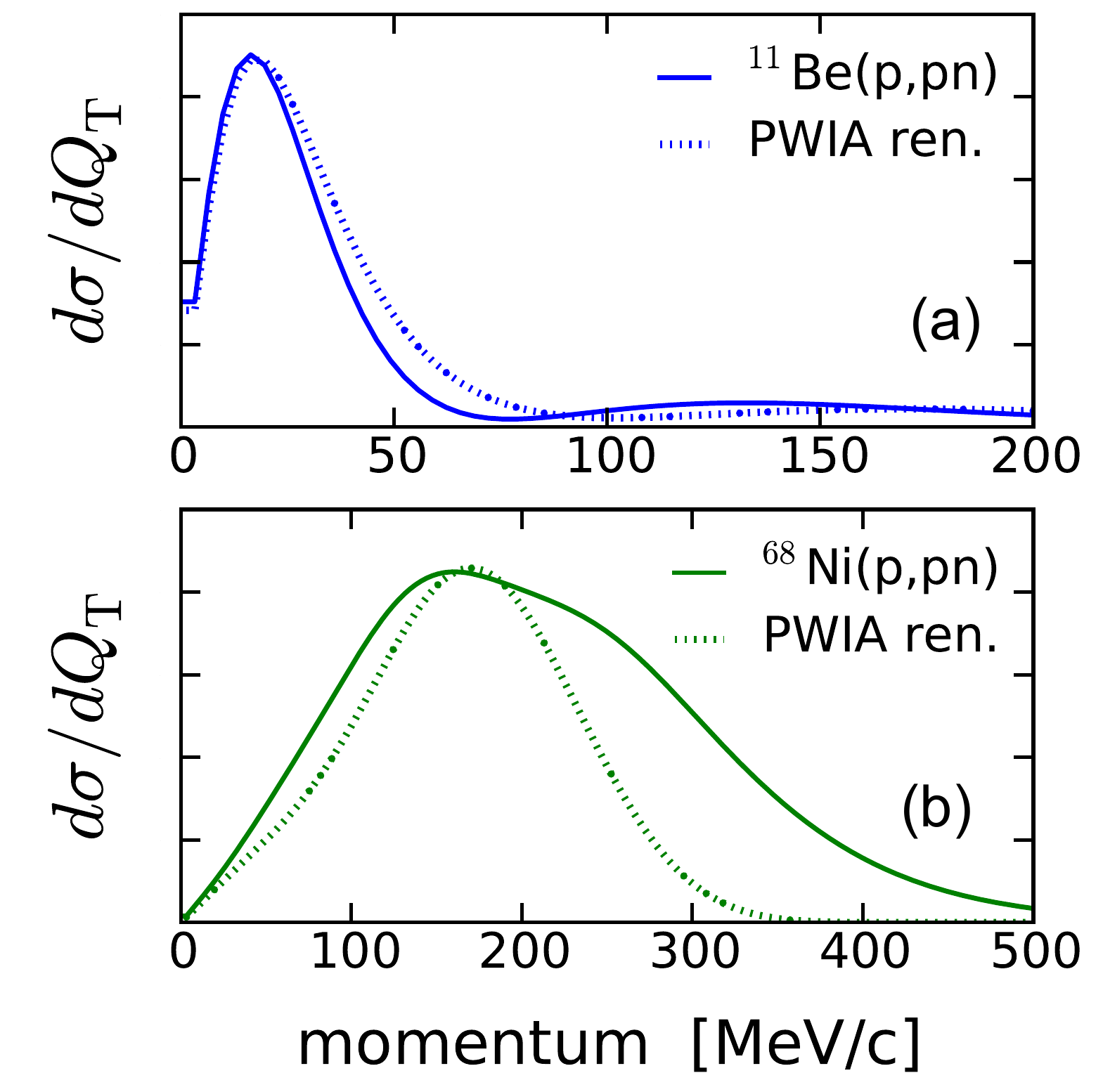}%
\caption{(Color online). Same as in figure \ref{widthqz}, but for transverse momentum distributions.}%
\label{widthqt}%
\end{center}
\end{figure}

Finally, in figures \ref{widthqz} and \ref{widthqt} we show the longitudinal (figure \ref{widthqz}) and transverse (figure \ref{widthqt}) momentum distributions for $^{11}$Be(p,n) and and $^{68}$Ni(p,n) reactions at 500 MeV/nucleon. For $^{11}$Be  we assume neutron removal from the 1s$_{1/2}$ state with 0.502 MeV separation energy, while for $^{68}$Ni we assume the removal from 0f$_{7/2}$ orbital with separation energy of 15.68 MeV.   In the Plane Wave Impulse Approximation (PWIA), the S-matrices for the incoming and outgoing particles are all set to the unity (no distortion). From Eq. \eqref{momdis} we get 

\begin{eqnarray}
\left( {d\sigma\over d ^3  Q}\right)_{PWIA} &=&  {1 \over (2\pi)^3} {S(lj)\over 2j+1} \sum_m \left< {d\sigma_{pN}\over d\Omega} \right>_Q \nonumber\\
&\times& \left| \int d^3 {\bf r} \ e^{-i{\bf Q.r}}
\psi_{jlm}({\bf r}) \right|^2 . \label{momdispwia}
\end{eqnarray}
In figures \ref{widthqz} and \ref{widthqt} the PWIA results are shown as dashed lines, whereas the solid lines are for the full DWIA, Eq. \eqref{momdis}. As we have discussed in connection with figure \ref{probf}, the PWIA  yields results that are much larger than in the DWIA, because the absorption is missing in the PWIA method. Hence, we renormalize the results to be shown in the same plot. It is clear from figure \ref{widthqz} that the shape of the longitudinal momentum distributions  is insensitive to the details of the wave function being probed. The two cases compared ($^{11}$Be and $^{68}$Ni) are very representative  because they differ strongly on the separation energies, angular momenta, and on  principal quantum numbers (number of nodes in the wave functions). In both cases, the  insensitivity of the width is manifest.   The physical reason is the same as for longitudinal momentum distributions in  knockout reactions, first shown in Ref. \cite{BM92}. This is easily understood from Eq. \eqref{momdisz} because the dependence on $Q_z$ is basically contained in the last integral: the S-matrices and in medium quasi-free cross sections are weakly dependent on $Q_z$. Their net effect is to rescale the magnitude of the total cross section.

For the transverse momentum distributions, we see a clear separation of shapes in figure \ref{widthqt}, as they are appreciably influenced by the S-matrices which enter directly into the $b$-integral in Eq. \eqref{momdist}. Some of the features of the wave distortion in the entrance and outgoing channels carry a stamp on the transverse momentum distributions, as seen in the upper panel of the figure where a shoulder in the tail of the distributions is clearly seen. This is the same sort of reaction distortion mechanism occurring in heavy ion knockout reactions which was emphasized in Ref. \cite{BH04}. The contribution of the different $m$ magnetic substates also impact on the shape of the distributions.  The DWIA and PWIA  results look  different, mostly visible as a shift of the curve for $^{11}$Be(p,pn), or strongly distorted, as in the case of $^{68}$Ni(p,pn) seen in the lower panel of figure   \ref{widthqt}.

\section{Conclusions}

There are other observables which have been used for nuclear spectroscopy with (p,pN) reactions. In the past, angular distribution of the nucleons have been scrutinized by comparison with theory. The comparison in this case is more sensitive to several details of the theory model. It has been indeed a difficult task to learn from experiments which   physical input was determinant for a good reproduction of the angular data \cite{JM73}. We believe that a much simpler method can be used based on momentum distributions, which also have been shown to be very useful in the case of heavy ion reactions.

One also probes the spectroscopy of nuclei more closely if one measure both (p,2p) and (p,pn) reactions and studies cross section ratios such as
\begin{equation}
{\cal R} = {(d\sigma/dQ)_{(p,pn)} \left(d \sigma_{pp}^{elast}/d\Omega\right)_{\theta = 90^\circ}     \over (d\sigma/dQ)_{(p,pn)} \left(d \sigma_{pn}^{elast}/d\Omega\right)_{\theta = 90^\circ} } .
\label{ratione}
\end{equation}
In fact, similar methods have been introduced in the past to separate the effects of final state interactions and off-shell effects in (p,2p) and (p,pn) cross sections at $E_p < 100$ MeV \cite{Wol93,Sho98,Sho00,Sho01}. 

In this paper we have focused on the opportunities that (p,2p) and (p,pn) reactions can offer to the studies of nuclei far from the stability line in experiments using inverse kinematics. Quasi-free (p,2p) and (p,pn) reactions have been used as a spectroscopic tool for more than 60 years. The reaction mechanism is known to be well described in the Distorted Wave Impulse Approximation formalism. We have developed a formalism making use of the eikonal theory allowing a quantitative description of the reaction mechanism including absorption from the elastic channel due to multiple-scattering effects. Our approach provides scattering-angle averaged cross sections and recoil momentum distributions, well adapted to the needs of large-acceptance experiments in inverse kinematics with radioactive beams. 

Here we concentrated on the use of momentum distributions as a spectroscopic tool, following the success of using nucleon knockout reactions for nuclear spectroscopy of radioactive beams. In fact, we have shown that (p,PN) reactions show some similarities with heavy-ion knockout reactions. But some striking differences are also  found. Perhaps the most evident one is that fact that (p,pN) reactions  are more sensitive to the interior part of the nuclear wave function. This carries imprints in the momentum distributions that are easily understood by using an eikonal and DWIA reaction formalism.  In view of the development and better experimental detection techniques, data are now of much higher quality than those obtained in previous decades. Pioneering experiments with (p,pN) with radioactive nuclei in inverse kinematics are now becoming available. In this article we have shown the main reaction mechanism features which are expected, and what can be learned from the properties of single particle states. More exclusive measurements will require more detailed reaction theory, which is now under development.  

%\bigskip 

\section{Acknowledgments}

We acknowledge beneficial discussions with Valerii Panin and Stephanos Paschalis. This work was partially supported by the US- DOE grants DE-FG02-08ER41533 and DE-FG02-10ER41706. One of the authors (C.B.) acknowledges the Helmholtz International Center for FAIR (HIC for FAIR) and the ExtreMe Matter Institute EMMI for supporting his visits to the Technische Universit\"at Darmstadt, where much of this work was done.

\end{document}